\documentclass[lettersize,journal]{IEEEtran}
\usepackage{amsmath,amsfonts}
\usepackage{array}
\usepackage[caption=false,font=normalsize,labelfont=sf,textfont=sf]{subfig}
\usepackage{textcomp}
\usepackage{stfloats}
\usepackage{url}
\usepackage{verbatim}
\usepackage{graphicx}
\usepackage{cite}
\usepackage{pifont}
\usepackage[]{subfig}
\usepackage{amsmath}
\usepackage{amsthm}

\usepackage{booktabs}
\usepackage{multirow,makecell}
\usepackage{tabularx}
\usepackage{booktabs}
\usepackage{threeparttable}
\usepackage[linesnumbered,ruled]{algorithm2e}

\begin{document}

\title{D-ADD: An Effective Plug-In for Defending Against Model Stealing}

\author{
    Jian-Ping Mei\IEEEauthorrefmark{1}\thanks{Corresponding author: Jian-Ping Mei.}, 
    Weibin Zhang,
    Jie Chen, 
    Xuyun Zhang, 
    Tiantian Zhu
    \thanks{
This work was supported by the National Natural Science Foundation of China (Grant No.62276234, U22B2028, and 62002324), and
Zhejiang Provincial Natural Science Foundation (Grant No:
ZCLZ24F0202).}
    \thanks{
    Jian-Ping. Mei, Weibin Zhang, Jie Chen, and  Tiantian Zhu were with the College of Computer Science and Technology,
Zhejiang University of Technology, Hangzhou 310023, China. E-mail: \{jpmei,ttzhu\}@zjut.edu.cn; zhangweibin30@gmail.com; chenjie.plus@qq.com.
}
\thanks{Xuyun Zhang was with the School of Computing, Macquarie University, 4 Research Park Drive, Macquarie Park, NSW 2109. E-mail: xuyun.zhang@mq.edu.au}
    }


\markboth{Journal of \LaTeX\ Class Files,~Vol.~14, No.~8, August~2021}%
{Shell \MakeLowercase{\textit{et al.}}: A Sample Article Using IEEEtran.cls for IEEE Journals}

\IEEEpubid{0000--0000/00\$00.00~\copyright~2021 IEEE}

\maketitle

\begin{abstract}
Malicious users attempt to replicate commercial models functionally at low cost by training a clone model with query responses. Timely prevention of such model-stealing attacks is challenging, as it requires achieving robust protection, maintaining utility, and ensuring low deployment overhead at the same time. In this paper, we propose a novel non-parametric detector called Account-aware Distribution Discrepancy (ADD) to recognize queries from malicious users by leveraging account-wise local query dependency. We formulate each class as a Multivariate Normal distribution (MVN) in the feature space and measure the malicious score as the sum of weighted class-wise distribution discrepancy. Together with shift adjustment, the ADD$^+$ detector is enhanced with the capability to manage domain shifts. When integrated with random-based prediction poisoning, it serves as a practical plug-and-play defense module, termed D-ADD, for image classification models. Results of extensive experimental studies show that D-ADD achieves strong defense against different types of attacks with little interference in serving benign users for both soft and hard-label settings. Codes are available from \url{https://github.com/AI-EXP-group/D-ADD}.
\end{abstract}


\section{Introduction}
\begin{figure}
	\centering
 	\includegraphics[width=1.0 \columnwidth]{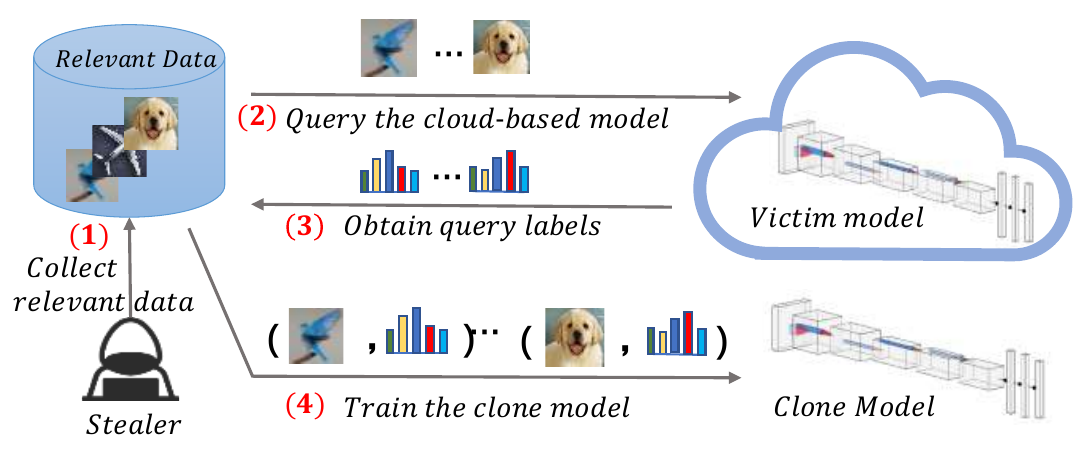}
 	\caption{Basic steps for stealing image classification models deployed on the cloud. By learning from the labeled queries, the final goal of the model stealer is to obtain a clone model with comparable classification accuracy at a low cost.}
	\label{fig.background}
 \end{figure}

\IEEEPARstart{D}{eep} neural networks that are trained on large-scale datasets have become essential tools in many machine learning tasks, especially in image classification ~\cite{krizhevsky2017imagenet,kolesnikov2020big,redmon2018yolov3}, where they have achieved state-of-the-art performance. However, training a powerful neural network model with a large number of parameters requires significant data and computation resources. As a result, companies that own the intellectual property of these models offer cloud-based services through APIs to generate profits. For instance, Google Cloud Vision API $\footnote{{https://cloud.google.com/vision/docs/labels}}$  provides a chargeable service for retrieving image category information through API queries. 
Unfortunately, when providing such services, the models to be queried are exposed to the risk of theft. Competitors or malicious users may be able to clone a model with comparable performance at a low cost and use it as a substitute for the target model functionally~\cite{DFME_truong2021,orekondy2019knockoff,KD_barbalau2020,DFMS_sanyal2022,tramer2016stealing}. This is known as ``model stealing''. A stealer makes use of the labeling service to annotate a set of query images, and use them to train their own models, as illustrated in Fig.\ref{fig.background}.
Model stealing could infringe on the commercial profit of the original service providers and even bring unfair market competition. It also poses threats to confidentiality as well as functionality if the stolen model is used as an auxiliary tool to carry out other inferential or adversarial attacks~\cite{FGSM}. Protecting models against
functionality cloning thus receives growing attention along the
research line of model-related privacy and security.

To mitigate the risk of model theft, labeling services are provided in a \underline{\textit{black-box}} manner, meaning that
visitors are not provided with any internal details of the model, intermediate results, or its training process. The only information that is returned in response to each query is a single class label (i.e., hard label) or a probability distribution over all the classes (i.e., soft label). By leaking out less clone-useful information in each response, the hard label setting helps to reduce the risk of being stolen compared to soft labels. Malicious users are assumed to have no or very limited in-distribution data. Model extraction becomes unnecessary when adversaries have access to adequate samples from the target model's training distribution, as they could then train a comparable model through alternative means. Under the \underline{\textit{data-limited}} condition, attackers resort to querying with 
surrogate samples or model generated examples.  Despite the fundamental protection provided by the black box and hard-label setting, it is still vulnerable to various model extraction attacks. It  has reported in~\cite{DFMS_sanyal2022} that even in the hard label and black-box setting, their attack can achieve about 93\%  performance of the target models on CIFAR-10 and CIFAR-100, and this number can reach up to
98\% for the simple MNIST task
\IEEEpubidadjcol

Various approaches have been proposed for defending against model stealing, but their limited defense capability and high deployment costs restrict them from practical use. Some approaches indiscriminately trigger poisoning through noise injection without checking whether a query is malicious or not~\cite{lee2018defending,PP_OrekondySF20,APMSA@TIFS23}. This ``always-poisoning'' strategy has to be applied with constrained degree of poisoning to maintain service utility for benign users, resulting in a limited defense.
A well performing malicious detector is thus a prerequisite for enabling strong defense.
Most existing detection-based defense approaches evaluate each query individually to predict whether it is ``malicious" or ``benign'' ~\cite{kariyappa2020defending,kariyappa2021protecting,CIP@TIFS23}.
However, malicious query detection is challenging for practical situations where some queries from benign users appear to be malicious while some attack queries are benign-looking. Moreover, learning a good discrimination ability often requires to collect enough malicious samples of various attacks. 

In this paper, we propose a new method to detect queries from malicious users using a technique called Account-aware Distribution Discrepancy (ADD). 
Our method measures the discrepancy between the query and training distributions using window-augmented feature statistics. Compared to the preliminary ADD detector proposed in \cite{D-ADD}, here we propose to integrate a shift adjustment component to enhance the detector's capability to handle global distribution shift of benign queries and name this variant as ADD$^+$.
By combining the ADD$^+$ detector with random-based prediction poisoning, we have a powerful defense module named D-ADD with the following advantages:

\begin{itemize}
	\item \textbf{A wide range of applications.} It provides strong defense and high utility against different types of attacks for both soft and hard label settings.
	\item \textbf{Low cost.} It is totally training-free and requires no collection of known malicious data. 
	\item \textbf{Deployment convenience.} It works in a favorable plug-and-play manner, and causes negligible response latency.
\end{itemize} 
We have conducted extensive empirical evaluations with various settings to verify its detection performance, defense capability for protecting image classification models from model stealing, as well as robustness to adaptive attack. Experimental results demonstrate that the proposed maliciousness detector substantially outperforms existing methods, even under challenging conditions involving both covariate and prior shifts. This enables D-ADD to provide significantly stronger protection against sophisticated cloning attacks compared to current defense mechanisms, while well preserving its original utility.

For the rest of this paper, we first briefly review existing attacking and defense approaches for black-box model stealing in Section \ref{sec:related} and then provide the threat model in Section \ref{sec:threat model} and further discuss the motivation of this study
in Section \ref{sec:motivation}. We then present the details of the proposed approach in Section \ref{sec:proposed}, followed by three groups of experimental studies. Finally, we conclude this work in Section \ref{sec:conclusion}.

\section{Related Work \label{sec:related}}
\subsection{Black-Box Model Stealing Attacks} 
Black-box model stealing refers to a technique where malicious users attempt to functionally copy a target model without having any knowledge or access to its weights, architecture, training data, or intermediate results.
A stolen or clone model is sometime created to assist with black-box adversarial example generation~\cite{CloudLeak20}. These adversarial attacks start with a small number of seed images from the target model's training dataset, and then synthesize examples by iteratively adding imperceptible perturbations using techniques such as JBDA ~\cite{papernot2017practical} and its variants~\cite{juuti2019prada}. Since the direct purpose of this type of attacks is to approximate the decision boundaries of the target model instead of functional replication, models cloned with adversarial attacks have much lower classification accuracy compared to other attacks ~\cite{papernot2017practical}.
In this paper, we consider model-stealing with the goal of creating a low-cost functionality clone of a target image classification model. 

Model-stealing attackers typically have little or no access to the proprietary training data of the target model. To handle this data-limited situation, attackers use surrogate or model-generated data as queries.
Using substitute samples as queries, the KnockoffNets attack comes to be a strong baseline if the surrogate dataset has high semantic or distributional similarity to the training data~\cite{orekondy2019knockoff}. Other studies borrow ideas from data-free knowledge distillation to steal with queries synthesized by maximizing prediction confidence~\cite{KD_barbalau2020} or the output discrepancy between the target and clone models ~\cite{MAZE2021,DFME_truong2021}. The DFMS attack in ~\cite{DFMS_sanyal2022}  has been shown to achieve high clone model performance with hard labels.

\subsection{Defense Against Black-Box Model Stealing}
Defense approaches can be broadly categorized into two groups based on whether they perform maliciousness evaluation.
Some approaches apply perturbation injection indiscriminately to poison any returned response~\cite{lee2018defending, WangWH24}. Perturbations are learned based on certain objectives, such as maximizing the clone model's training loss or restricting predictions near the decision boundary, and can be injected into the final predictions ~\cite{PP_OrekondySF20,Mazeika0F22}, inputs ~\cite{APMSA@TIFS23}, or outputs of intermediate layers ~\cite{LeeH022PertuFeatureMap}. These approaches aim to reduce knowledge leakage from the returned label without significantly impacting the utility for benign users. Although later approaches like Prediction Poisoning (PP) ~\cite{PP_OrekondySF20}  can tolerate higher levels of perturbations by sacrificing some level of utility, poisoning must still be limited due to their indiscriminate nature, which leads to minimal protection especially for attacks working with hard labels. 

For defense approaches that perform poisoning only to suspicious queries, the identification of queries from malicious users becomes critical. Many of these approaches evaluate each query individually to predict whether it is ``malicious'' or ``benign'' using a detector learned from known malicious queries ~\cite{kariyappa2020defending,kariyappa2021protecting,CIP@TIFS23}. 
However, the final performance of these approaches is not guaranteed when the malicious query set they learn from is limited in scale and coverage due to the high collection cost.

Some other studies proposed defense methods for specific types of model extraction attacks.  For example, stateful detection approaches such as PRADA are particularly tailored for the detection of adversarial example generation, where the query sequence consists of images with high self-similarity visually~\cite{chen2020stateful, juuti2019prada}, and the retraining based defense works only for data-free model extraction attack~\cite{defenseDFME23}.

\section{Threat Model} \label{sec:threat model}

\textbf{Black-box labeling service.}
The API-based labeling service with a high-performance image classification model is provided to registered users in a black-box manner.  The defender is aware of the account information that each query is submitted. The target model consists of an encoder $g_{\boldsymbol{\theta}}\left(\cdot\right)$ and a classifier head $f_{\boldsymbol{\theta}}\left(\cdot\right)$. For any given query image $\mathbf{x}$, the model returns a class label $\hat{\mathbf{y}}=g_{\boldsymbol{\phi}}(f_{\boldsymbol{\theta}}(\mathbf{x}))$, i.e., a vector $\mathbf{y}$  of length $K$. The label is said to be ``soft'' if it contains probabilities of all the $K$ classes, and ``hard'' if it is one-hot. 

\textbf{Goals of the attacker and the defender.}
The attackers' goal is to train a clone model with comparable classification performance to the target model using label information returned by querying the target model. The
defender's goal is to create a defense mechanism that stops leaking positive knowledge learned by the service model to malicious users, or specifically, minimizing the performance of the stolen model while having little impact on normal user experience.

\textbf{Data availability and Assumption.}
It is typical that users have no access to the original training data of the target model as the data itself may be the most important property that its owner want to protect. 
In most existing studies \cite{kariyappa2020defending,kariyappa2021protecting,WangWH24}, it is commonly assumed that queries from malicious users are out-of-distribution (OOD) samples.
In contrast, rather than enforcing every query to be an OOD sample, we adopt a weaker assumption to presume that there exists a distributional separation between benign and malicious queries. Formally, we state it as following, i.e.,
\underline{\textit{Assumption}}: queries from benign users are overall with larger semantic similarities to the training data than those from malicious users, and the target encoder is able to capture the semantic gap. This assumption is essential to ensure the solvability of the detection problem.

\section{Motivation} \label{sec:motivation}
Despite various defense approaches being proposed,
the challenge of providing effective and deploy-friendly protection against the latest functionality-oriented model stealing attacks remains unresolved. Current defense approaches, including ``active'' approaches that add perturbations to each query and ``post-detection'' ones that only perform poisoning upon detection exhibit notable limitations as listed in Table \ref{tb:exisiting}. These limitations are summarized as follows.

\begin{itemize}
\item Limited defense capability. While selective poisoning allows more substantial prediction manipulation compared to those ``always poisoning'' approaches, its ultimate efficacy critically depends on the capability of malicious detection. 
The detector's performance is inherently constrained by the representativeness and quality of the training sample set, from which it learns to differentiate benign and malicious queries.

\item Restricted application scope. Most approaches are only applicable to the soft-label setting. Active defense offers almost no protection against hard-label-based stealing methods since accuracy-preserved prediction perturbation can be easily removed by keeping only the highest probability.

\item Large cost and high complexity. 
Active defenses employing optimization-based noise injection require complex training regimens, including multi-loss balancing \cite{APMSA@TIFS23} and surrogate network-assisted gradient estimation \cite{PP_OrekondySF20}, incurring high implementation complexity. 
 On the other hand, detection-based defense approaches need to retrain the already deployed models together with the detector to gain the detection ability, causing a large overhead and creating deployment bottlenecks due to tight model-detector coupling.

\item Lack of mechanism for dealing with domain shift. Domain shifts are a prevalent challenge in real-world deployment scenarios. These methods typically assume queries from benign users strictly conform to the training data distribution and lack explicit mechanisms to handle distributional discrepancies. This limitation undermines their practical utility.
\end{itemize}

\begin{table*}
	\centering
	\caption{ Limitations of query-wise defense approaches}\label{tb:exisiting}
	\begin{tabular}{cccccc}
		\toprule
		 Method   & detector &requirement of malicious/OOD data& applicable to hard-label & re-training/fine-tuning  & dealing with domain shift\\ \midrule
		DP\cite{lee2018defending}         & \ding{55} & \ding{55} & \ding{55} & \ding{51} & \ding{55} \\
		MAD\cite{PP_OrekondySF20}         & \ding{55} & \ding{55} & \ding{55} & \ding{51} & \ding{55} \\ 
        GRAD$^2$\cite{Mazeika0F22}        & \ding{55} & \ding{55} & \ding{55} & \ding{55} & \ding{55}\\
		APMSA\cite{APMSA@TIFS23}          & \ding{55} & \ding{55} & \ding{55} & \ding{51} & \ding{55} \\
        MeCo\cite{defenseDFME23}          & \ding{55} & \ding{55} & \ding{51} & \ding{51} & \ding{55} \\
        ACT\cite{WangWH24}                & implicit  & synthetic & \ding{51} & \ding{51} & \ding{55} \\
	    EDM\cite{kariyappa2021protecting} & \ding{51} & \ding{51} & \ding{51} & \ding{51} & \ding{55} \\
		AM\cite{kariyappa2020defending}   & \ding{51} & \ding{51} & \ding{55} & \ding{51} & \ding{55} \\		
		CIP\cite{CIP@TIFS23}              & \ding{51} & \ding{51} & \ding{51} & \ding{51} & \ding{55} \\
		PRADA\cite{juuti2019prada}        & \ding{51} & \ding{55} & \ding{51} & \ding{55} & \ding{55} \\

		\bottomrule
	\end{tabular}	
\end{table*}
\textbf{The potential of non-parametric detectors.}
As we discussed previously, detecting malicious queries is crucial for maintaining a strong defense. Although detectors with learnable parameters can be effective, they require additional training and rely heavily on pre-collected malicious data. Non-parametric approaches are potential alternatives, especially when detection is performed without the use of malicious data. One straightforward method is to directly apply non-parametric out-of-distribution (OOD) detection metrics, such as confidence ~\cite{hendrycks2016baseline} or energy~\cite{EnergyOODNIPS20}, to identify malicious queries by treating in-distribution (ID) samples as benign queries and OOD samples as malicious ones.
Instead of evaluating each query individually, stateful approaches treat the queries from each account entirely. One such approach called PRADA evaluates maliciousness at the account level by analyzing whether queries submitted by a certain account follow a normal distribution ~\cite{juuti2019prada}. PRADA calculates the Shapiro-Wilk score  ~\cite{shapiro1965analysis}  with the distances between a stream of consecutive queries to measure how closely these queries follow a normal distribution. It raises an alert when the score falls below a specified threshold. 

\textbf{Limitation of PRADA.}
In PRADA, it is assumed that queries submitted by any malicious account come from different distributions with significant distances. This assumption fits well with adversarial example generation attacks, where queries combine seed images from the original training distribution and synthesized samples from another distribution.
PRADA works effectively in detecting adversarial-oriented types of attacks such as JBDA~\cite{papernot2017practical}. However, it may not be able to detect surrogate queries used by KnockoffNets~\cite{orekondy2019knockoff} when surrogate samples are from the same distribution.

\begin{figure*}[!t]
	\centering
	\subfloat[\scriptsize{Steal MNIST classifier with FashionMNIST samples.}]{
		\includegraphics[width=0.23\linewidth, height =0.18\linewidth]{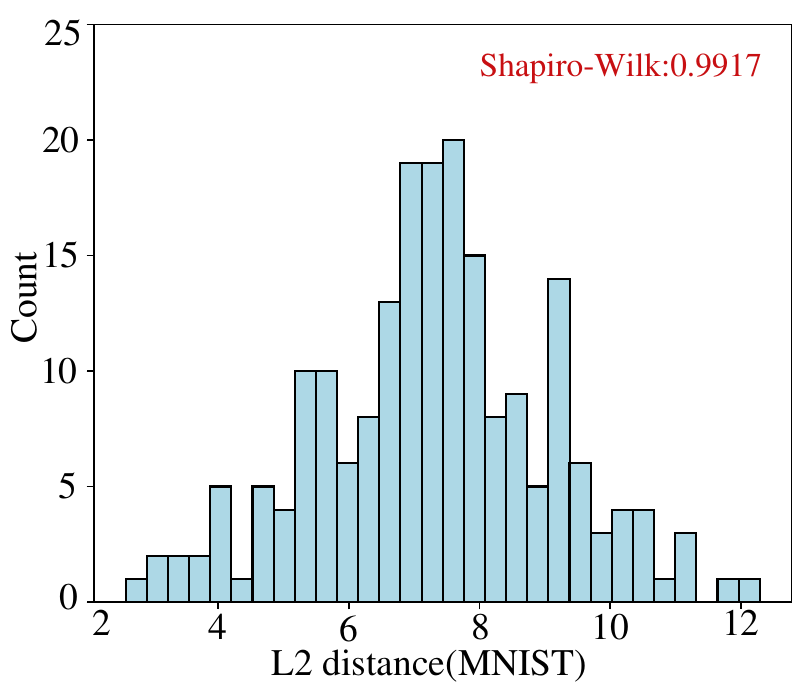} 
		\includegraphics[width=0.23\linewidth, height =0.18\linewidth]{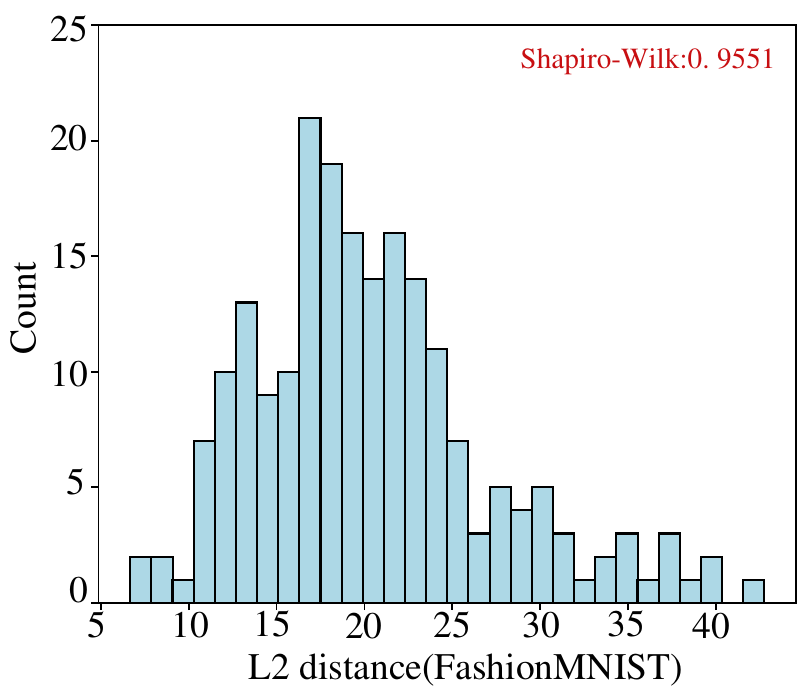}
	}
	\hfil
	\subfloat[\scriptsize{Steal CIFAR-10 classifier with CIFAR-100 samples.}]{
		\includegraphics[width=0.23\linewidth, height =0.18\linewidth]{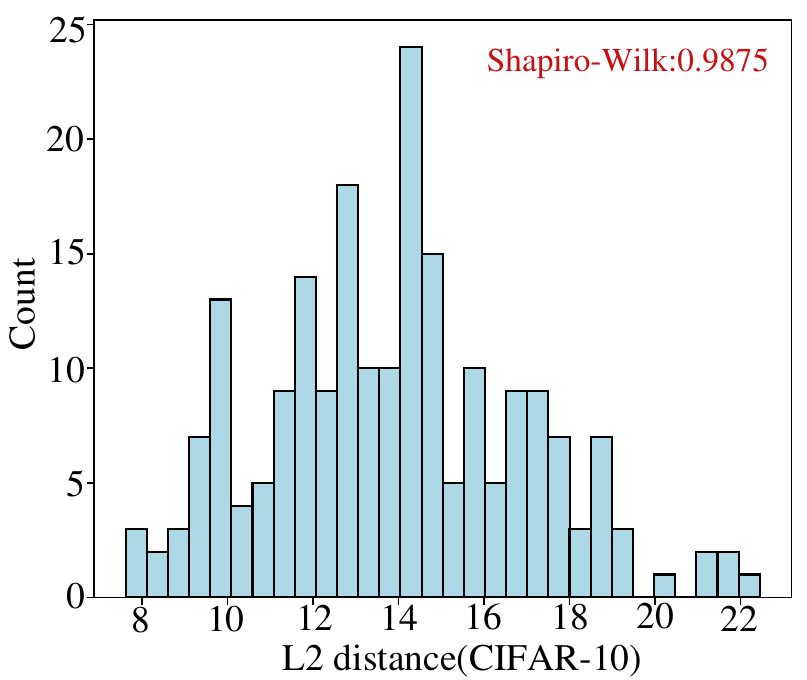}
		\includegraphics[width=0.23\linewidth, height =0.18\linewidth]{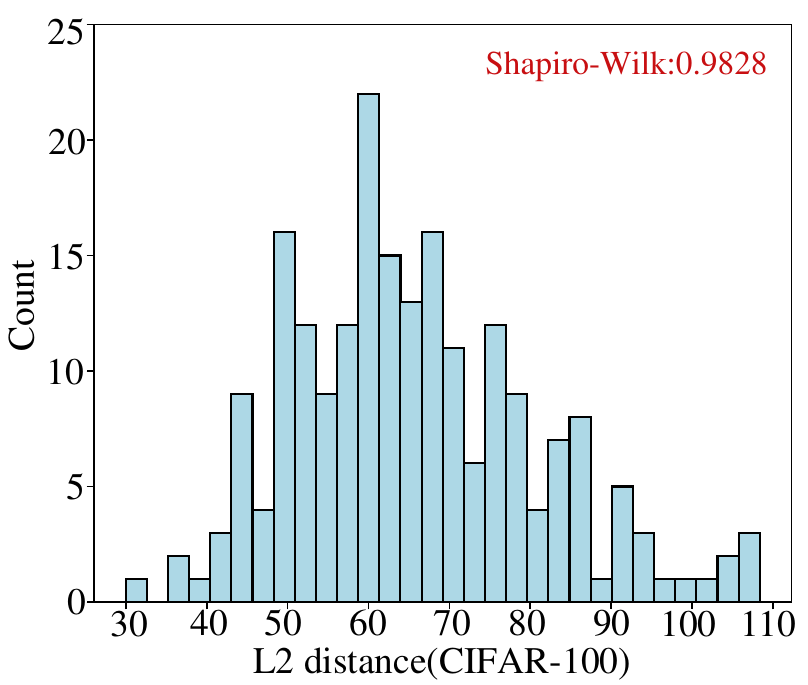}			}
	\caption{Comparison of the normality of distribution of distances produced with PRADA between \textbf{benign} queries (on the \textbf{left}) and \textbf{malicious} queries (on the \textbf{right}).  The KnockoffNets stealing attack is simulated by using FashionMNIST and CIFAR-100 to query the classification model trained on MNIST and CIFAR-10, respectively. The Shapiro-Wilk score ranged in $[0, 1]$ measures the fitness of a set of values to a normal distribution.}
	\label{prada.toy} 
	
\end{figure*}

\begin{figure*}[]
	\centering
	\subfloat{
		\centering
		\includegraphics[width=0.3\linewidth, height =0.22\linewidth]{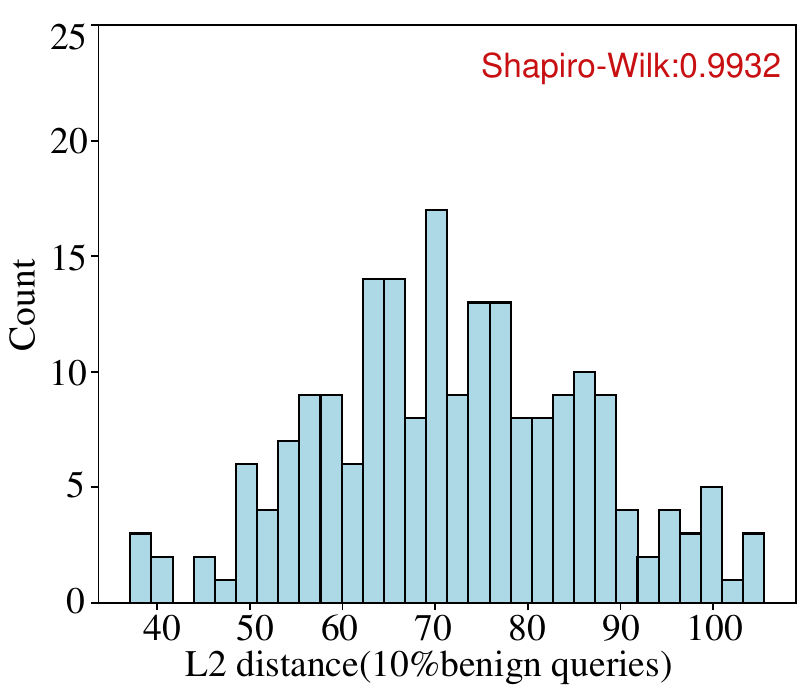} 		\hfill
		\centering
		\includegraphics[width=0.3\linewidth, height =0.22\linewidth]{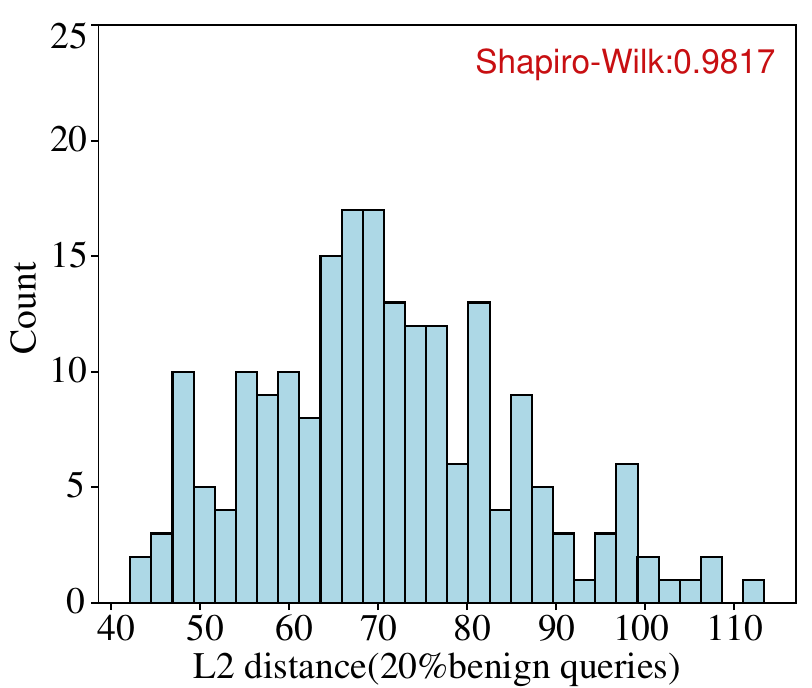}\hfill
		\centering
		\includegraphics[width=0.3\linewidth, height =0.22\linewidth]{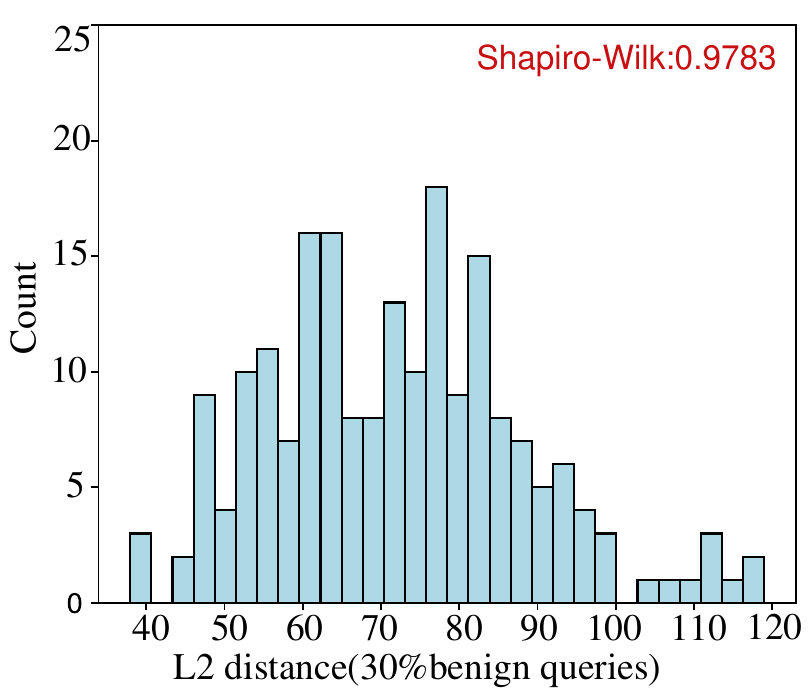}
		\label{prada_mix}
	}
	\caption{Histograms produced by PRADA with mixed surrogate/OOD (CIFAR-100) and in-distribution (CIFAR-10) images. The  in-distribution percentage is 10\%, 20\% and 30\%, respectively.}
	\label{prada.mix} 
\end{figure*}

To provide empirical evidence, we plotted the detection results of PRADA in Fig. \ref{prada.toy}  when a malicious account attempted to steal MNIST and CIFAR-10 classification models using FashionMNIST and CIFAR-100 samples as surrogate queries, respectively. This figure shows that it is difficult to distinguish malicious queries from benign ones from the shape of their histogram. Although the Shapiro-Wilk scores of malicious histograms are smaller than the benign ones, the gap is subtle. We get similar results when histograms are calculated with queries containing both surrogate and in-distribution samples in Figure \ref{prada.mix}, showing that PRADA is quite insensitive to the distribution difference of natural image queries even when the mixing ratio is as large as 30\%. This insensitivity makes PRADA unsuitable for handling generic types of functionality cloning attacks including KnockoffNets~\cite{orekondy2019knockoff}, DFME~\cite{DFME_truong2021} and DFMS~\cite{DFMS_sanyal2022}.

\section{Defense with Account-aware Distribution Distance (D-ADD) \label{sec:proposed}} 

Although non-parametric detectors are attractive for their simplicity in concept as well as deployment, existing ones are not effective enough to provide robust protection against various stealing attacks. Malicious query detection faces significant practical challenges due to the inherent overlap between benign and adversarial query distributions. Notably, certain legitimate user queries may exhibit even greater deviation from training examples than actual attack queries, leading conventional query-wise detectors to generate excessive false alarms.


\begin{figure*}
	\centering
	\includegraphics[width=0.85\textwidth, height=0.35\textwidth]{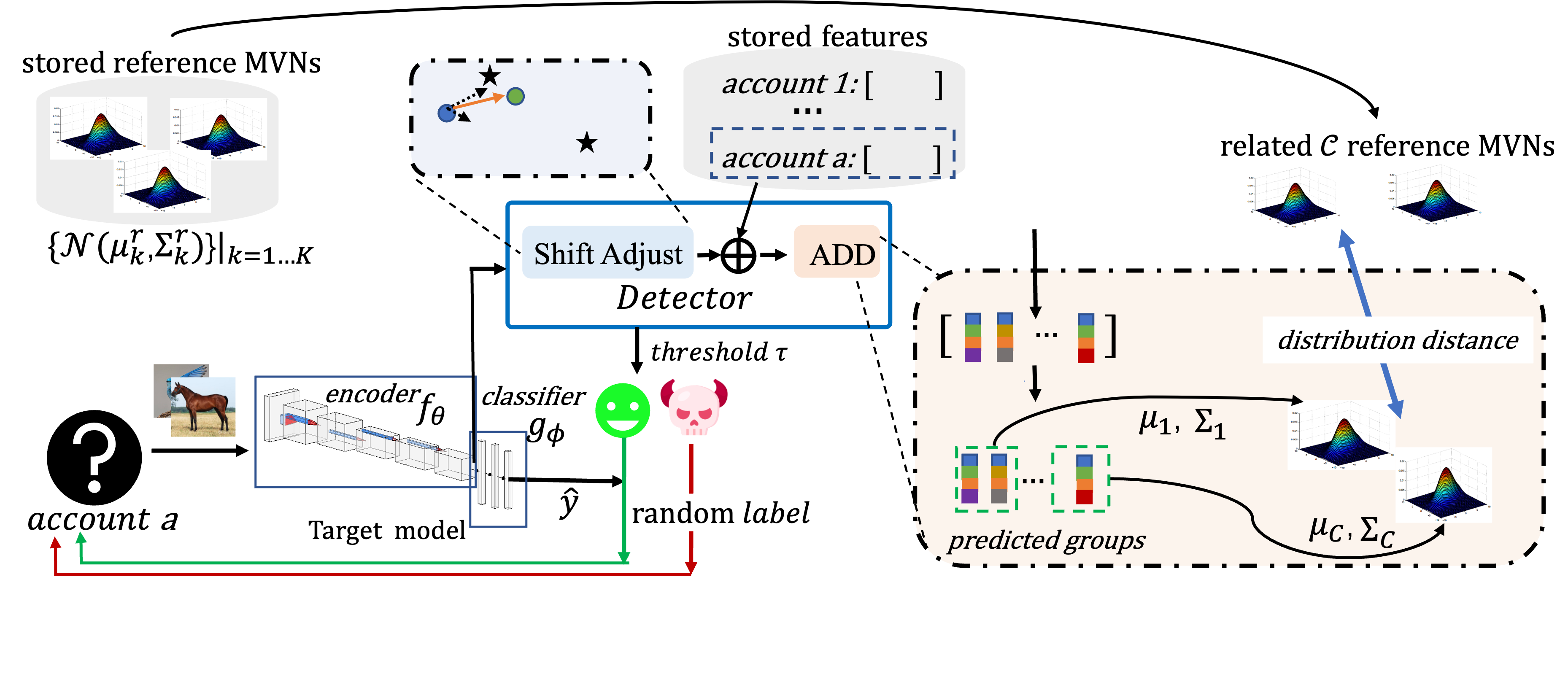}
	\caption{Overall pipe line of the proposed D-ADD Defense. The the new embedding space detector consists of two components: shift adjust and ADD for handling domain shift and calculating the account-based distribution distance.}
	\vspace{-0.1in} 
	\label{fig2.toy}
\end{figure*}

In this paper, we further exploit the potential of detection-based defense by proposing
a new method called D-ADD. It comprises two main components: an Account-aware Distribution Distance (ADD) detector and a random-based prediction poisoning step, as illustrated in Fig. \ref{fig2.toy}. A randomly selected label is returned to mislead clone model training if the query is found to be from a malicious user. The decoupling formulation with the target model allows for seamless integration and removal of this defense module.
Since the poisoning strategy used here is quite straightforward, we focus on the embedding space malicious detector, which first performs feature adjustment of a submitted query and then calculates a malicious score based on ADD. To distinguish it from  the vanilla ADD detector, we designate the detector with domain shift adaptation as ADD-plus, denoted ADD$^+$.

The overall idea of ADD is to measure the maliciousness of query according to how much the query distribution deviates from the training distribution in the output space of the target encoder.
Unlike stateful approaches in~\cite{chen2020stateful} and PRADA ~\cite{juuti2019prada} that establish global dependency among all queries from the same account, we only establish local dependency between nearby queries in the sequence and measure the discrepancy between two distributions based on feature statistics. We will discuss the merits of these designs in Section \ref{sec:discussion}.
Next, we present the detailed formulation of these two components one by one.

\subsection{ADD-based malicious score}

As sketched on the right of Fig. \ref{fig2.toy}, the main idea of ADD is to measure a shift-adjusted query distribution with the reference data distribution in the embedding space. However, directly formulating the query or training distribution is challenging. 
Drawing inspiration from data-free knowledge distillation ~\cite{choi2020data}, we assume that feature statistics of each class follow a Multi-Variate Normal (MVN) distribution $\mathcal{N}(\boldsymbol{\mu},\boldsymbol{\Sigma})$.

To get a reference distribution of each class, we store the mean vector and covariance matrix for each of the $K$ classes  based on features of training samples in that class. A sliding window is allocated and maintained for each account to store the adjusted features as well as the final predictions of a number of latest queries. The stored features of the corresponding account are retrieved to enlarge the coming queries to obtain more reliable statistical information. 

Given the window-augmented query batch $Q$, we partition them into $C$ groups according to their predicted classes with the largest probability. Let $Q=\mathcal{X}_1\bigcup\mathcal{X}_2\dots\mathcal{X}_C$, where the subset $\mathcal{X}_c$ contains features of queries that predicted to be in the $c$th class.
We then calculate the mean vector $\boldsymbol{\mu}^a_c$ and covariance matrix $\boldsymbol{\Sigma}^a_c$ with $\mathcal{X}_c$, and measure the class-wise  discrepancy based on the distance between the two distributions of this class, i.e.,
\begin{align}
	d_c(\boldsymbol{\mu}^r_c,\boldsymbol{\Sigma}^r_c, \boldsymbol{\mu}^a_c,\boldsymbol{\Sigma}^a_c)= dist(\mathcal{N}^r_c(\boldsymbol{\mu}^r_c,\boldsymbol{\Sigma}^r_c),\mathcal{N}^a_c(\boldsymbol{\mu}^a_c,\boldsymbol{\Sigma}^a_c))  \nonumber \\
	=\Vert\boldsymbol{\mu}^r_c-\boldsymbol{\mu}^a_c\Vert^{2}+\operatorname{tr}\left[\boldsymbol{\Sigma}^r_c\!+\!\boldsymbol{\Sigma}_c^a\!-\!2\left(\boldsymbol{\Sigma}^r_c \boldsymbol{\Sigma}^a_c\right)^{\frac{1}{2}}\right],
	\label{eq:EachclassFD}
\end{align}
where $dist(\mathcal{N}^r_c, \mathcal{N}^a_c)$ denotes the squared Fréchet distance between the reference and the query-batch distribution of class $c$~\cite{dowson1982frechet}.

\begin{figure}
	\centering
	\includegraphics[width=0.95 \columnwidth]{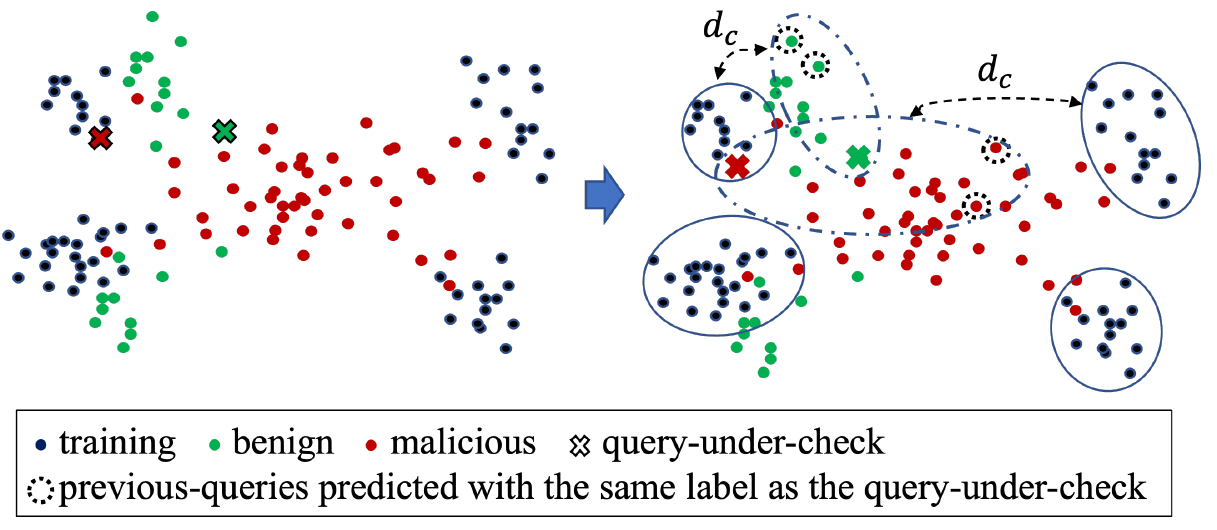}
	\caption{Illustration of the working principle of ADD. With
the class-wise discrepancy $d_c$, ADD recognizes the benign
query (green cross) and the malicious query (red cross), even
the malicious one is closer to some of the training samples.}
	\label{fig1.toy}
\end{figure}

The Malicious Score (MS) for the query batch $Q$ is then calculated as		
the sum of weighted class-wise distance over all the $C$ classes, i.e.,
\begin{align}
	MS_{ADD}(Q) =\sum_{c\in \mathcal{C}} \frac{|\mathcal{X}_c|}{N} d_c(\boldsymbol{\mu}^r_c,\boldsymbol{\Sigma}^r_c, \boldsymbol{\mu}^a_c,\boldsymbol{\Sigma}^a_c). \label{eq:maliciousScore}
\end{align}
The larger the MS, the more possible that the current query batch is from a malicious user.
Here each batch may only involve a subset of the $K$ classes, i.e., $C \leq K$. Fig.\ref{fig1.toy} illustrates the ability of ADD to detect benign-looking malicious queries and benign-looking queries.

\subsection{Feature adjustment for dealing with domain shift}

The formulation of class-wise distance in Eq.(\ref{eq:maliciousScore}) helps ADD to deal with prior shift or label distribution inconsistency between benign query distribution and training distribution (see discussion in Section \ref{sec:simplifiedVersion}). This ensures our detector to maintain good performance when benign queries only involve a small subset of classes (See results in Table \ref{tb:in-distribution} and Fig. \ref{fig:ROC}). However, covariate shift also commonly exists in real application due to changes in background, resolution, and so on. Relying solely on the domain generalization ability of the encoder itself may not enough to ensure good detection performance in some complex circumstances given a limited window size. 

To enhance the robustness of the ADD detector with respect to global distribution shift in the embedding space, we propose a gravity-based query feature adjustment. Specifically,
 for sample $\mathbf{x}_i$, we adjust its original feature $f_{org}(\mathbf{x}_i)$ to obtain a new one with
\begin{align} 
f_{new}(\mathbf{x}_i) &=f_{org}(\mathbf{x}_i)+\alpha\sum_{c=1}^K \hat{y}_{ic} \vec{\Delta}_{ic}, \label{eq:adjust}\\
\text{with} \quad \vec{\Delta}_{ic}&=\hat{y}_{ic} (\boldsymbol{\mu}^r_c - f_{org})
\end{align}
Here, $\hat{y}_{ic}$ represents the predicted probability of sample $x_i$ to class $k$, $\boldsymbol{\mu}^r_c$ is the mean vector of class $c$, and $\alpha \in (0, 1)$ is the adjustment coefficient. 
$\vec{\Delta}_{ic}$ represents the attraction from the center of the $c$th class to this sample, and gravity in this direction is further weighted by the predicted probability of this class $\hat{y}_{ic}$. 
The overall adjustment for each query feature is decided by the attraction from all the class centers. 

\begin{figure}[!t]
	\centering
	\subfloat[before adjustment]{
        \includegraphics[width=0.5\linewidth, height=0.38\linewidth]{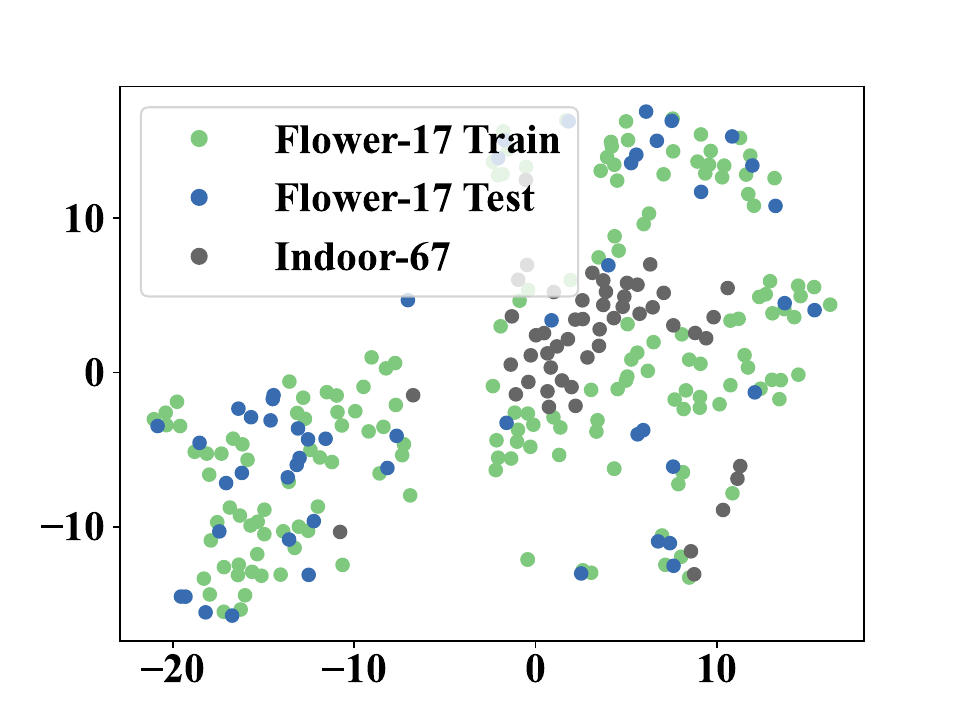}
        \label{}
    }
	\subfloat[after adjustment]{
        \includegraphics[width=0.5\linewidth, height=0.38\linewidth]{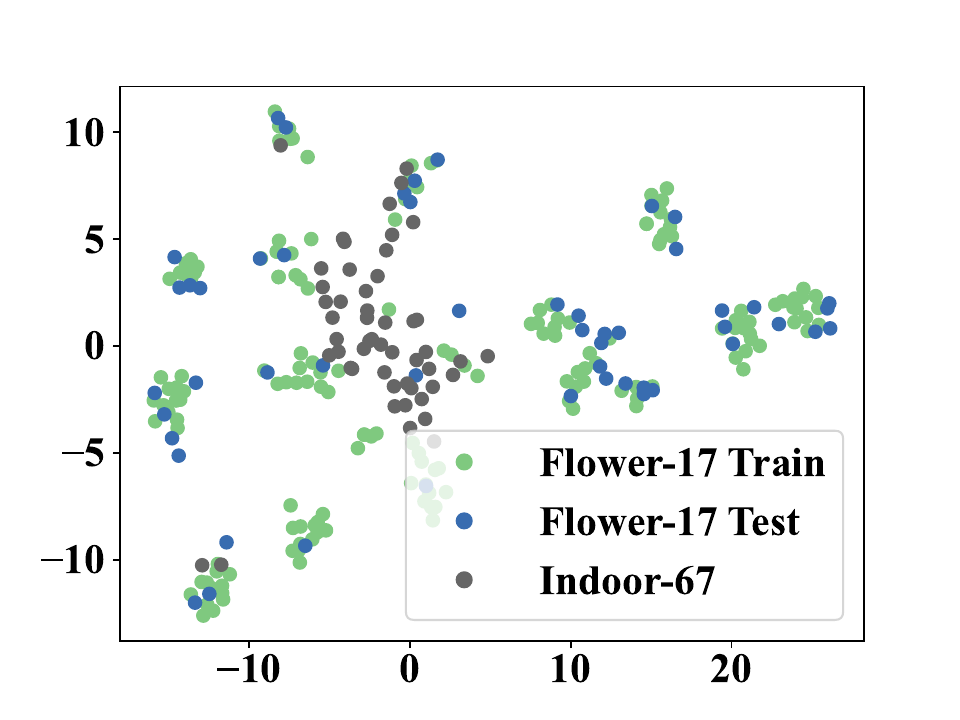}   
        \label{}
    }
	\caption{Feature space query distributions before and after shift-adjustment ($\alpha=0.5$) for the Flower-17 model. We compare the adjustment effect on in-distribution (ID) query (Test samples of Flower-17) and out-of-distribution (OOD) query (samples from Indoor-67). Results are visualized with t-SNE.}
	\label{fig:tsne} 
\end{figure}

 Fig.~\ref{fig:tsne} illustrates the effect of the proposed feature adjustment on in-distribution samples and OOD samples with a target model trained on Flower-17. Here in-distribution queries are samples from the test set of Flower-17 and OOD queries are samples from another dataset Indoor-67. By comparing the query feature distribution before and after adjustment, it is seen that samples being much closer to one particular class center form a more compact cluster around that class center due to the large attraction from that direction. On the contrary, samples being far from all the class centers as most of the OOD samples almost remain unadjusted because of small and balanced ``gravitational forces'' from all the classes. Such kind of adjustment is beneficial for dealing with the covariate shift by reducing the global distribution distance between benign queries and training ones. As shown by our experimental results (See Fig.\ref{fig.alpha}), our detector requires a much smaller window size to achieve comparable detection performance by applying feature adjustment. Note that feature adjustment is applied only within the detector, leaving the classification layer input unchanged.

\subsection{Implementation}
\textbf{Main steps.} Algorithm \ref{alg:D-ADD} gives the main steps of the proposed D-ADD defense. The reference distributions of all the classes are obtained once the target model is trained and remains unchanged afterwards. The detection module first apply shift-adjustment to the feature of the coming queries and then calculates the means and the covariance matrices with the updated window for calculating the squared Fréchet distance. By comparing this distance to a properly chosen threshold, we can readily determine whether a query is malicious or not.  Once a query is recognized as malicious.

\begin{algorithm}
	\DontPrintSemicolon
	\SetAlgoLined
	\KwIn {Trained target modal with encoder $f_{\boldsymbol{\theta}}\left(\cdot\right)$ and classifier $g_{\boldsymbol{\phi}}\left(\cdot\right)$, statistics of of each reference class $\{(\boldsymbol{\mu}^r_c,\boldsymbol{\Sigma}^r_c)\}_{c=1}^K$, feature window of size $N$ for each account, query sample $ \mathbf{q}$ currently submitted by account $a$, threshold $\tau$, coefficient $\alpha$.}
	\KwOut {Class label for the current query $\mathbf{q}$.}
	
	
    
	get prediction $\hat{\mathbf{y}}=g_{\boldsymbol{\phi}}(f_{\theta}\left(\mathbf{q}  \right))$ for $\mathbf{q}$
    
    get adjusted query feature $f_{new}(\mathbf{q})$ using  Eq.(\ref{eq:adjust}) with $f_{org}(\mathbf{q})=f_{\theta}\left(\mathbf{q}\right)$

    update the window of account $a$ with $(f_{new}(\mathbf{q}),\hat{\mathbf{y}})$ to get $Q$ 
    
   calculate $\boldsymbol{\mu}^a_c,\boldsymbol{\Sigma}^a_c$  for each involved class of $Q$
	
calculate the malicious score $MS_{ADD}$ with Eq.(\ref{eq:maliciousScore}) and Eq.(\ref{eq:EachclassFD});

	\tcp{random-based poisoning}
	\uIf{$MS_{ADD} > \tau$}{
		\Return{a random soft or hard label}
	}
	\Else{
		\Return{ $\hat{\mathbf{y}}$ or $\arg_c\max \hat{y}_c$ }
	}
	
	\caption{D-ADD} \label{alg:D-ADD}
	
\end{algorithm}

\textbf{Sliding window.}
For each account, we maintain a window of size $N$ to store the encoder-extracted features as well as the final predictions of the latest queries. When an account $a$ submits $S$ images at a time for labeling, the feature vectors of these images are adjusted and pushed into its window to form an updated batch of features for calculating the feature statistics  of each involved classes.
Initially, the window is filled with $N$ random samples from the target model's training set to ensure that the target model preserves utility during the startup phase.

\textbf{Detector threshold.\label{sec:tau_setting}}
Like other non-parametric detectors, we need to set an appropriate threshold $\tau$ to produce the final detection in the proposed defense module. To provide a feasible way to control the trade-off between utility and safety, we propose setting $\tau$ based on a specified tolerance level of performance degradation, which is measured by a dropping ratio $\gamma$ in the testing accuracy of the D-ADD integrated target model. 
The testing accuracy increases gradually with the increase of $\tau$, resulting in reduced impact of the defense module. When the threshold reaches a specific level, we obtain the maximum testing accuracy $acc^*$, which is the same as the original undefended model. We can then manage the service quality of the model after defense by specifying $\gamma$. 
As the preserved testing accuracy does not necessarily lead to the true utility, we may trade off less testing accuracy by setting $\tau$ to be slightly larger than the one that leads to the testing accuracy of ${acc^*}\times (1-\gamma)$.

\textbf{Complexity and storage overhead.}
ADD is training-free. Given window size $N$ and embedding dimension $d$, the time complexity of straightforward implementation for calculating a covariance matrix is $O(Nd^2)$ and multiplication of two covariance matrices in the  Fréchet distance is $O(d^3)$.  Since both $N$ and $d$ are typically less than one thousand, the time complexity is small. In our experiment, the response latency caused by defense is less than 5ms per query.

Low-dimensional features instead of original images are stored. The storage overhead required to allocate each account a small window is negligible for modern cloud-based servers. For instance, storing  $64$ features, each of which is  a 256-dimensional floating-point vector would require 64KB. This means that it is possible to accommodate up to 16,384 users simultaneously with 1GB of storage space.

\section{Discussions} \label{sec:discussion}
\subsection{Advantages over simplified variants}\label{sec:simplifiedVersion}

In our default formulation in Eq.(\ref{eq:maliciousScore}), the malicious score of a given query batch is the sum of weighted class-wise distances of feature statistics.  For each involved class of the query batch, a distance is calculated to measure its deviation to the corresponding reference class, and is weighted by the size of this class. Next, we consider two possible variants.

\textbf{Equal Weights (EW).}
Each class is equally weighted when calculating the total distance, i.e., 
\begin{align}
	MS_{EW}(Q)=\sum_{c\in \mathcal{C}}  d_c(\boldsymbol{\mu}^r_c,\boldsymbol{\Sigma}^r_c, \boldsymbol{\mu}^a_c,\boldsymbol{\Sigma}^a_c). 
\end{align}

\textbf{Global Distribution Distance (GDD).} The feature statistics are calculated based on the global distribution without considering each class individually, i.e.,
\begin{align}
	MS_{GDD}(Q)=d(\boldsymbol{\mu}^r,\boldsymbol{\Sigma}^r, \boldsymbol{\mu}^a,\boldsymbol{\Sigma}^a)
\end{align}
where $\boldsymbol{\mu}^r$ and $\boldsymbol{\Sigma}^r$ are the mean and covariance matrix calculated based on all the training samples, and $\boldsymbol{\mu}^a$ and $\boldsymbol{\Sigma}^a$ are calculated based on all query features in $Q$.

As evidenced by experimental results (See Fig.\ref{fig:ROC}), both class-conditional and weight by class size are crucial for enabling ADD to handle more challenging situations, where queries from benign users are not strictly in-distribution but exhibit various types of distribution shifts.

\subsection{Impact of shift-adjustment}
In order to provide another point of view, we rewrite the above formula in Eq.(\ref{eq:adjust}) into a new form as below
\begin{equation} 
f_{{new}}(\mathbf{x}_i) = (1 - \alpha) f_{{org}}(\mathbf{x}_i) + \alpha \sum_{c=1}^K \hat{y}_{ic} \boldsymbol{\mu}^r_c.
\end{equation}
The above formulation shows that the new feature is a convex combination of the original feature and the weighted global mean $\sum_{c=1}^K \hat{y}_{ic} \boldsymbol{\mu}^r_c$.
When $\alpha \rightarrow 0$, it leads to no adjustment. When $\alpha \rightarrow 1$, the feature is directly updated to the weighted global mean. Selecting an appropriate value for $\alpha$ is crucial to control the degree of feature adjustment, as excessive adjustment may bring a negative impact to the detector performance. We found setting $\alpha \in (0.3, 0.7)$ generally produce reasonable results in various cases.

We also provide the matrix form below for efficiently applying feature adjustment when $B$ queries come at a time, i.e., 
\begin{equation} 
\mathbf{F}_{{new}} = (1 - \alpha) \mathbf{F}_{{org}} + \alpha \hat{\mathbf{Y}} \cdot \mathbf{M},
\end{equation}
where  \( \mathbf{F}_{\text{org}} \in \mathbb{R}^{B \times d} \) and  \( \mathbf{F}_{\text{new}} \in \mathbb{R}^{B \times d} \) are matrices containing the original and the adjusted $d$ dimensional features of $B$ queries, \( \hat{\mathbf{Y}} \in \mathbb{R}^{B \times K} \) is the matrix of predicted probabilities, and \( \mathbf{M}\in \mathbb{R}^{K \times d} \) is the matrix containing $K$ class mean vectors. 

\subsection{Robustness to adaptive attack \label{sec:defineRobust}}
Attackers can try to evade a defense technique once they understand its working principle. 
In our case, the attacker may manipulate the query stream by mixing benign-looking samples into the window to tweak the batch-based query distribution towards a benign one so that the whole batch will be incorrectly labeled as benign. While combing more benign-looking samples makes it easier to evade ADD, the attacker is assumed to have very limited access to such kind of data.

\textbf{ Smaller window leads to more robustness.}
Assume that the attacker has $H$ benign-looking (BL) samples and abundant malicious-looking (ML) samples, and has the knowledge of the minimum percentage of BL in each window required to evade ADD is $x$\%, but unaware of the window size. We consider the attack query sequence with a length of $H\times(1+M)$ is organized with each single BL sample followed by $M$ ML samples, where $M=\lceil 100/x \rceil -1$.  Noting that this is the maximum value of $M$ that ensures the overall percentage of BL queries in the sequence is at least $x\%$. The uniform mixing-up of BL and ML throughout the sequence leads to the largest number of subsequences with a BL percentage not less than $x\%$. 
 
Based on the above attack sequence, we can derive that the number of malicious queries including BL and ML that missed by ADD is related to window size $N$, i.e.,
\[
no.missed=
\begin{cases} 
&(\overbrace{1}^{BL}+\overbrace{N-1}^{ML})\times H \   \ \text{for  } N \leq M\\
&(\overbrace{1}^{BL}+\overbrace{M}^{ML})\times H \   \ \text{for  } N \geq M +1
\end{cases}
\]

\textbf{Detection vs. robustness.} A large window gives more reliable statistics and we are even able to implicitly realize an infinitely large window ($N\to \infty$) by  incrementally updating the statistics. Nevertheless, a small window provides more sensitivity to changes in query sequences and leads to more robustness as shown in the above formula.

\subsection{Benefits of detection in embedding space} 
As a commercialized service provider, the target model is expected to have been trained to generalize to distribution shifts caused by input-space perturbations like changes of background, resolution, etc. This gives the following benefits by working as an embedding space detector. It allows ADD maintain good utility when queries from benign users are not strictly in-distribution. Moreover, when an attacker tries to evade ADD by increasing the percentage of benign-looking samples via duplication or augmentation, we can detect such behavior by observing  abnormally high occurrence of very similar features. This should work especially well for encoders trained by contrastive learning, which maps augmented inputs of the same image to be closer to each other than those of different images \cite{SimCLR-Chen20}.

\begin{table*}[t]
    \centering
    \caption{Comparison of different malicious detectors. The number in brackets is the window size $N$ used in ADD$^+$.}
    \label{tb:in-distribution}
    \renewcommand{\arraystretch}{1.2}
    \resizebox{\textwidth}{!}{
    \begin{tabular}{cccccccccccccc}
        
        \multicolumn{14}{l}{\textbf{Scenario-1: in-distribution benign queries}} \\
        \toprule
        \multirow{2}{*}{\textbf{Training Data}} & \multirow{2}{*}{\textbf{Benign vs. Malicious}} & \multicolumn{4}{c}{\textbf{FPR@TPR95 ($\downarrow$) (\%)}} & \multicolumn{4}{c}{\textbf{AUROC ($\uparrow$) (\%)}} & \multicolumn{4}{c}{\textbf{AUPR ($\uparrow$) (\%)}} \\
        \cmidrule(r){3-6} \cmidrule(r){7-10} \cmidrule{11-14} 
        & & \textbf{Baseline} & \textbf{Energy} & \textbf{PRADA} & \textbf{ADD$^+$($N$)} & \textbf{Baseline} & \textbf{Energy} & \textbf{PRADA} & \textbf{ADD$^+$($N$)} & \textbf{Baseline} & \textbf{Energy} & \textbf{PRADA} & \textbf{ADD$^+$($N$)} \\
        \hline
        
        MNIST     & MNIST vs. FashionMNIST    & 5.26  & 2.16  & 95.64 & \textbf{0.00} (16)    & 98.65 & 99.53 & 63.04 & \textbf{100.00} (16) & 99.71 & 99.91 & 88.26 & \textbf{100.00} (16) \\
        FashionMNIST & FashionMNIST vs. MNIST & 90.56 & 91.78 & 95.24 & \textbf{0.00} (16)    & 59.90 & 71.04 & 50.00 & \textbf{100.00} (16) & 89.76 & 94.18 & 50.00 & \textbf{100.00} (16) \\
        CIFAR-10  & CIFAR-10  vs. CIFAR-100   & 34.46 & 35.47 & 95.52 & \textbf{0.03} (16)    & 89.14 & 90.45 & 41.80 & \textbf{99.99} (16) & 96.84 & 97.42 & 79.46 & \textbf{99.99} (16) \\
        CIFAR-100 & CIFAR-100 vs. CIFAR-10    & 60.14 & 65.17 & 70.21 & \textbf{4.78} (32) & 76.79 & 77.19 & 85.52 & \textbf{98.84} (32)& 92.72 & 92.96 & 97.30 & \textbf{98.97} (32)\\
        Flower-17 & Flower-17 vs. Indoor-67   & 51.18 & 63.53 & 17.65 & \textbf{0.00} (32)   &84.41  & 87.57 &97.50  & \textbf{99.79} (32) &99.76  &99.82  & 99.05 &\textbf{99.98} (32)  \\
        
        \toprule
        \\
        
        \multicolumn{14}{l}{\textbf{Scenario-2: benign queries from distribution with shifts}} \\
        \toprule
        \multirow{2}{*}{\textbf{Training Data}} & \multirow{2}{*}{\textbf{Benign vs. Malicious}} & \multicolumn{4}{c}{\textbf{FPR@TPR95 ($\downarrow$) (\%)}} & \multicolumn{4}{c}{\textbf{AUROC ($\uparrow$) (\%)}} & \multicolumn{4}{c}{\textbf{AUPR ($\uparrow$) (\%)}} \\
       \cmidrule(r){3-6} \cmidrule(r){7-10} \cmidrule{11-14} 
        & & \textbf{Baseline} & \textbf{Energy} & \textbf{PRADA} & \textbf{ADD$^+$($N$)} & \textbf{Baseline} & \textbf{Energy} & \textbf{PRADA} & \textbf{ADD$^+$($N$)} & \textbf{Baseline} & \textbf{Energy} & \textbf{PRADA} & \textbf{ADD$^+$($N$)} \\
        \hline
        
                & USPS-10 vs. FashionMNIST & 56.81 & 53.08 & 100.00& \textbf{1.99} (16) & 81.89 & 83.11 & 34.84 & \textbf{99.45} (16) & 96.56 & 96.84 & 84.69 & \textbf{99.48} (16) \\
        MNIST   & USPS-7 vs. FashionMNIST  & 56.96 & 53.19 & 98.03 & \textbf{3.38} (16)& 81.82 & 83.06 & 56.90 & \textbf{99.27} (16) & 97.56 & 97.76 & 93.31 & \textbf{99.38} (16) \\
                & USPS-3 vs. FashionMNIST  & 57.58 & 53.66 & 25.11 & \textbf{5.83} (16) & 81.55 & 82.86 & 96.89 & \textbf{98.92} (16) & 98.94 & 99.03 & 99.82 & \textbf{99.58} (16) \\\midrule
                & STL-9 vs. CIFAR-100      & 57.80 & 58.11 & 49.20 & \textbf{5.69} (16) & 77.41 & 80.21 & 91.26 & \textbf{99.04} (16) & 96.56 & 97.14 & 99.19 & \textbf{99.10} (16) \\
        CIFAR-10& STL-7 vs. CIFAR-100      & 58.17 & 58.47 & 81.72 & \textbf{4.85} (16) & 77.21 & 80.04 & 69.41 & \textbf{99.12} (16) & 97.28 & 97.74 & 96.05 & \textbf{99.33} (16) \\
                & STL-3 vs. CIFAR-100      & 57.72 & 58.92 & 20.43 & \textbf{3.03} (16) & 76.84 & 79.85 & 97.50 & \textbf{99.38} (16) & 98.79 & 99.01 & 99.93 & \textbf{99.77} (16) \\
        
        \toprule
    \end{tabular}
    }
\end{table*}



\section{Experiments \label{sec:app:results}}

To evaluate the proposed approach, we present three groups of experiments. We first evaluate the performance of the ADD$^+$ detector, study the impact of window size, and also compare with two simplified variants in \ref{sec:exp_detection}. We then assess the effectiveness of our proposed defense approach in safeguarding neural classifiers against three state-of-the-art functionality clone attacks in \ref{sec:exp_stealing}. Finally, in Section \ref{sec:exp_adaptiveAttack}, we provide empirical results to show the capability of D-ADD in dealing with adaptive attacks.
All experiments were implemented using PyTorch 1.10 and conducted on a server equipped with an NVIDIA GeForce RTX 4090 GPU (24GB RAM), CUDA 11.2, and Ubuntu 20.04 LTS. Codes will be released once this paper is accepted. 

\textbf{Datasets.}
We trained five target models on MNIST, FashionMNIST, CIFAR-10 CIFAR-100, and Flower-17, and used another three datasets to simulate malicious or benign queries. Each involved datasets is summarized below.
\begin{itemize}
\item MNIST. This dataset consists of 70,000 grayscale images of handwritten digits (0-9), each of size 28$\times$28 pixels. These images are split into two sets: 60,000 images for training and 10,000 images for testing. Each image corresponds to one of the ten digit classes.
\item FashionMNIST. This dataset consists of 70,000 grayscale images of fashion-related items, designed as a more challenging alternative to the MNIST dataset. Each image is 28$\times$28 pixels in size and corresponds to one of 10 fashion classes. The dataset is split into a training set of 60,000 images and a test set of 10,000 images.
\item CIFAR-10 and CIFAR-100 \cite{cifar100}. Both datasets contains 60,000 32$\times$32 color images that are split into two subsets with 50,000 for training and 10,000 for testing.  CIFAR-10 consists of 10 classes of the same size, and CIFAR-100 is a more challenging variant of the CIFAR-10 that contains 100 equal-sized classes.  
\item FLower-17. This dataset contains 17 different species of flowers, with each class containing 80 images, resulting in a total of 1,360 images. The images vary in terms of size, orientation, lighting, and background.
\item Indoor-67. This dataset contains 15,620 images belonging to 67 different indoor scene categories. Each category has at least 100 images collected from public sources such as Flickr and Google, with varying viewpoints, lighting conditions, and occlusions. 
\item USPS. This dataset consists of 9,298 grayscale images of handwritten digits (0-9), each of size 16$\times$16 pixels. It is a classic benchmark for handwritten digit recognition tasks. Compared to the MNIST dataset, USPS images are smaller, noisier, and more challenging, making it useful for testing the robustness of classification models.
\item STL-10. This dataset contains 13,000 labeled images of size 96$\times$96, divided into 10 classes similar to CIFAR-10 but with higher resolution and more complex object variations. It also includes 100,000 unlabeled images for unsupervised and semi-supervised learning. Unlike CIFAR-10, which has 50,000 training samples, STL-10 only provides 500 training samples per class, making it a more challenging dataset for supervised learning.
\end{itemize}


\subsection{Experiments-I: Malicious Detection \label{sec:exp_detection}}

\textbf{Target models and malicious queries.}
Following~\cite{kariyappa2021protecting}, we have trained five target models on well-known image datasets, namely MNIST, FashionMNIST, CIFAR-10, CIFAR-100, and Flower-17 for image classification. We follow~\cite{kariyappa2020defending} to set model architectures and surrogate data to simulate malicious queries given in Table \ref{tab:architecture}. 


\textbf{Benign queries.}
We evaluate the performance of each detector in two scenarios.
In this first scenario, benign users have \underline{\textit{in-distribution}} queries.   For each target model, we use the corresponding testing set as benign queries.  
While this setting has been widely adopted in the literature, it is overly idealized since users typically do not have access to samples from the data distribution of the target model in real-world applications. Therefore, we further consider a more challenging scenario where benign queries exhibit \underline{\textit{distribution shifts}}.  It
occurs due to various perturbations in the input space, such as changes in background, corruption, and resolution. Distribution shift  is also attributed to label distribution inconsistency, for instance, when users query with images from only a subset of classes rather than all the classes that the target model is trained on. 
Results are reported based on MNIST and CIFAR-10 models that we found suitable benign queries.
For the MNIST model, we use samples from the USPS dataset  as benign queries. USPS also contains images of ten digits like MNIST but collected from a different source, therefore exhibiting feature distribution shifts. To further consider label inconsistency, we extract USPS-7 and USPS-3 to contain images of seven  or three  randomly sampled digits, respectively. For the CIFAR-10 model, we use samples from the STL-10 dataset as benign queries.  We discard the images of \textit{monkey} from STL-10 as it is not included in CIFAR-10, leaving nine classes that appeared in both datasets. Similarly, another two subsets denoted as STL-7 and STL-3 are also used to simulate label distribution inconsistency.

\textbf{Compared methods and Evaluation metrics.}
We compare the proposed ADD$^+$ (i.e., ADD with shift-adjustment) with the following three non-parametric detection approaches.
\begin{itemize}
	\item \textit{Baseline}~\cite{hendrycks2016baseline} checks the maliciousness of each query based on the maximum confidence, i.e., the highest probability of the soft label returned by the target model. Queries with low confidence are considered malicious.
	\item \textit{Energy}~\cite{EnergyOODNIPS20} detects malicious queries according to their energy values. Queries with high energy are regarded as malicious.
	\item \textit{PRADA}~\cite{juuti2019prada} detects malicious queries based on the normality of the distribution of distances between pairs of queries submitted from a certain account.
\end{itemize}

The detection is correct when a query from a \textit{benign account} is labeled as \textit{positive},  or a query from a \textit{malicious account} is labeled as \textit{negative}. 
Results are reported in terms of three widely used metrics, namely the area under the ROC curve (AUROC), the area under the precision-recall curve (AUPR), and False Positive Rate at 95\% True Positive Rate (FPR@TPR95), which indicates the percentage of negative samples that are mistakenly classified as positive ones given a True Positive Rate of 95\%.

\textbf{Comparison of detector performance.}
Table \ref{tb:in-distribution} gives the results of four detectors. The corresponding window size is given in brackets after the result of ADD$^+$. 
It is seen that ADD$^+$ performs consistently much better than other approaches in both scenarios. The performances of \textit{Baseline} and \textit{Energy} are comparable. Both of them perform well only for the simple MNIST model. The results of PRADA confirms that it often fails to work properly against a wide range of attacks due to its formulation limitation mentioned earlier. 

We found that ADD$^+$ correctly recognized all the malicious queries from benign ones given a sufficiently large window. The perfect detection indicates that we are able to separate all malicious queries from benign ones, or all benign query batches have smaller scores than malicious ones calculated with the proposed  measurement in Eq.(\ref{eq:maliciousScore}). Detailed results on the impact of window size are given later. A larger window size is needed to achieve good detection in more complex situations when the quality of  malicious queries are relatively high compared to benign ones. For example,  CIFAR-10 is used to steal the CIFAR-100 model. 	
Although the performance of ADD$^+$ degrades to some degree compared to in-distribution benign queries with the same window size for difficult benign queries,  it can still be remedied by using a larger window.  In practice, a 100\% detection rate may not be necessary, as a small number of false negatives may not provide sufficient information for effective model extraction.

\begin{figure*}[h]
		\centering
		\subfloat[\scriptsize{target model: MNIST}]{
			\includegraphics[width=0.31\linewidth,height=0.2\linewidth]{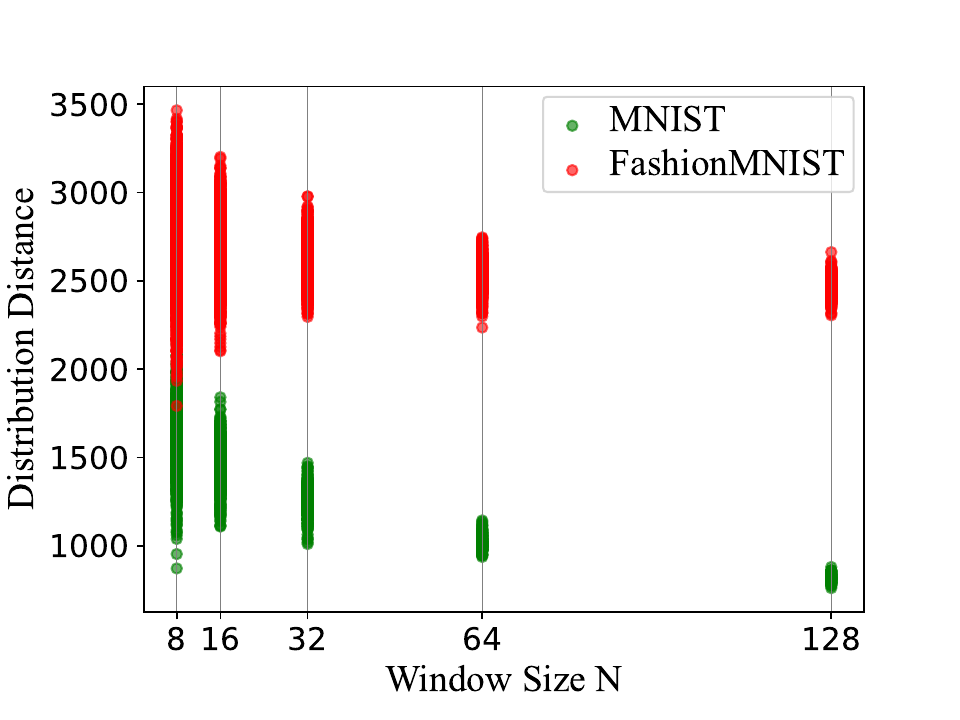}
		}
        \centering
	  \subfloat[\scriptsize{target model: FashionMNIST}]{
			\includegraphics[width=0.31\linewidth,height=0.2\linewidth]{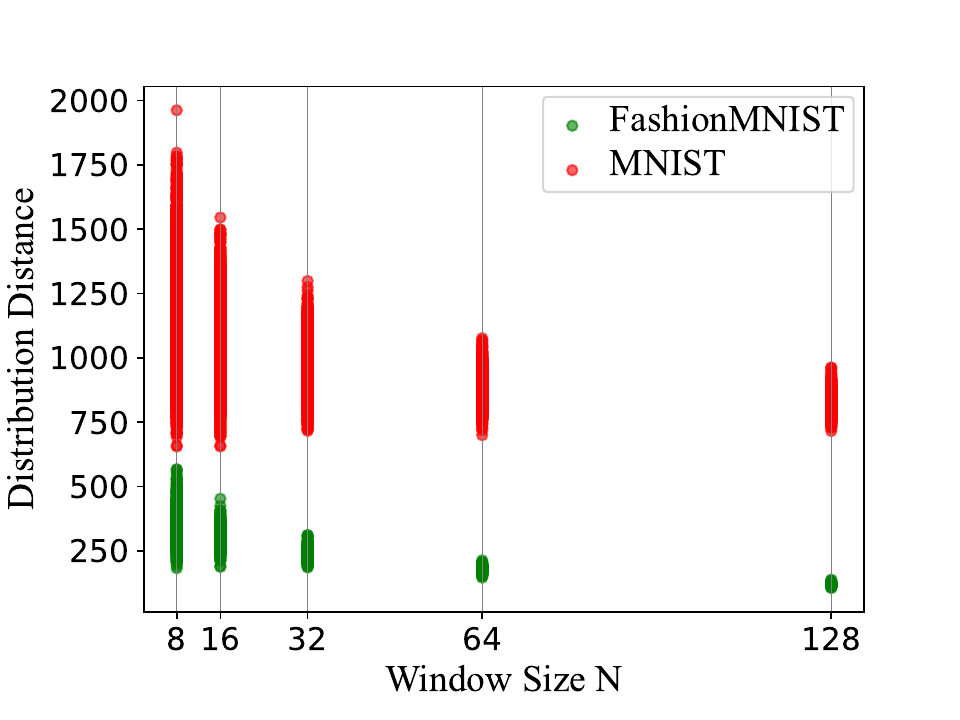} 	
		}
        \\
        \centering
	  \subfloat[\scriptsize{target model: CIFAR-10}]{
			\includegraphics[width=0.31\linewidth,height=0.2\linewidth]{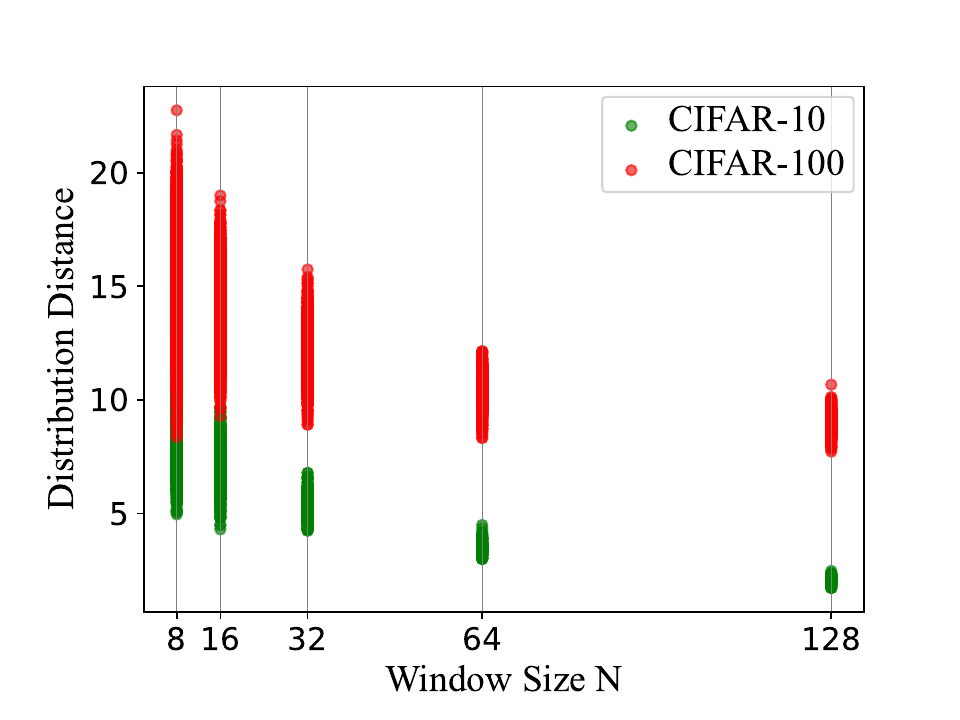}			
		}
        \centering
	  \subfloat[\scriptsize{target model: CIFAR-100}]{
			\includegraphics[width=0.31\linewidth,height=0.2\linewidth]{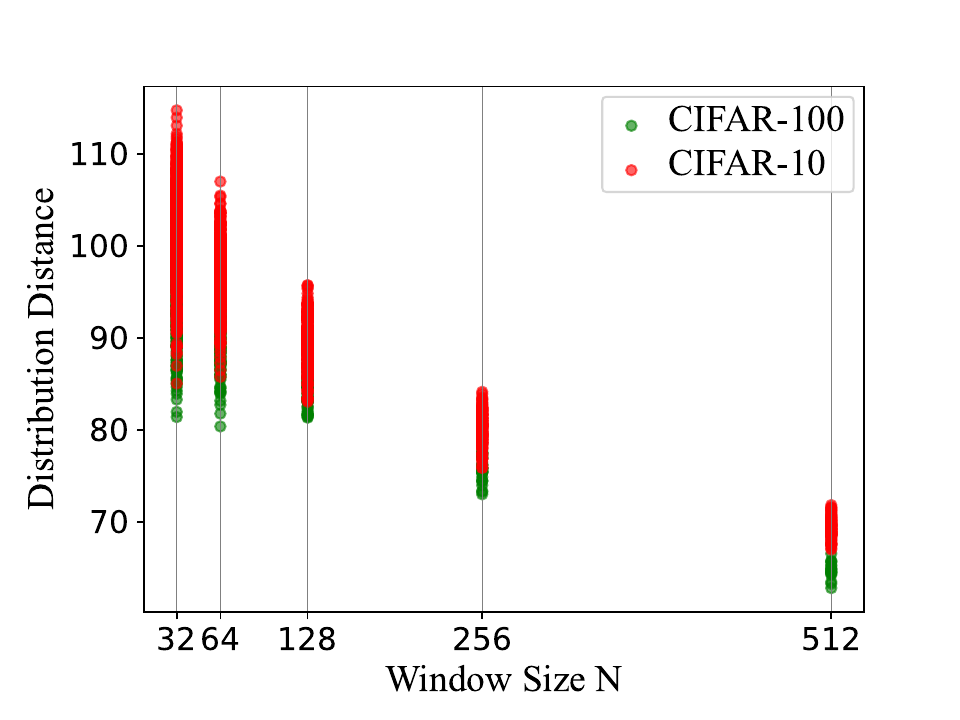}			
		} 
        \centering
	  \subfloat[\scriptsize{target model: Flower17}]{
			\includegraphics[width=0.31\linewidth,height=0.2\linewidth]{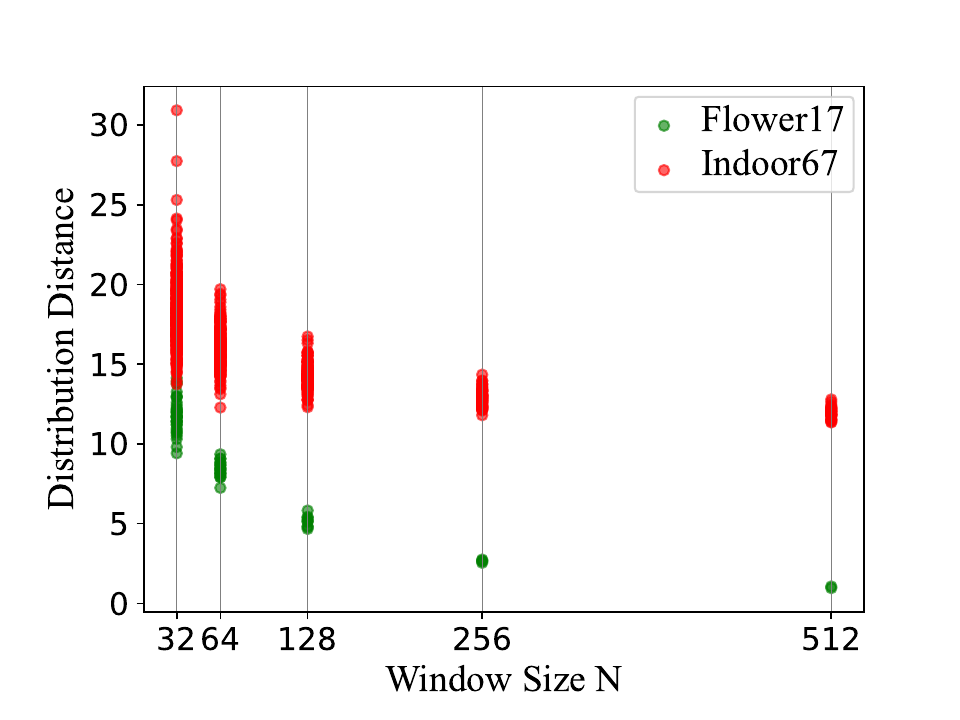}			
		} 
		\caption{Impact of sliding window size $N$ on distribution of Malicious Score (MS)s produced by ADD without feature adjustment. Each red dot is calculated with a window of $N$ randomly selected surrogate samples as malicious queries, and  each green dot is calculated with $N$ randomly selected testing images as benign queries.}
		\label{fig.Ksep}
\end{figure*}

\textbf{Impact of window size on detection.}
Now, we study the impact of window size $N$ on the separation of malicious scores between benign queries and malicious queries.  Here results are obtained with in-distribution benign queries. Conclusions are the same for benign queries with distribution shifts.  Fig. \ref{fig.Ksep} plots the malicious scores of benign query batches (green dots) and malicious query batches (red dots) with various window sizes. To focus on ADD itself, here we turn off shift-adjustment by setting $\alpha=0$. 

We observed that, a larger window size generally leads to a larger gap or a clearer separation between the scores of benign and malicious queries. 
While perfect detection can be achieved theoretically as long as no overlapping exists between the two sets of scores, a larger gap makes the selection of a proper threshold easier. However, for the more complex task of CIFAR-100, the degree of separation is still limited even we increased $N$ to 512. A large overlapping between those red and green dots indicates a large detection error caused by false positives and false negatives.

\begin{figure}
	\centering
 	\includegraphics[width=0.9 \columnwidth]{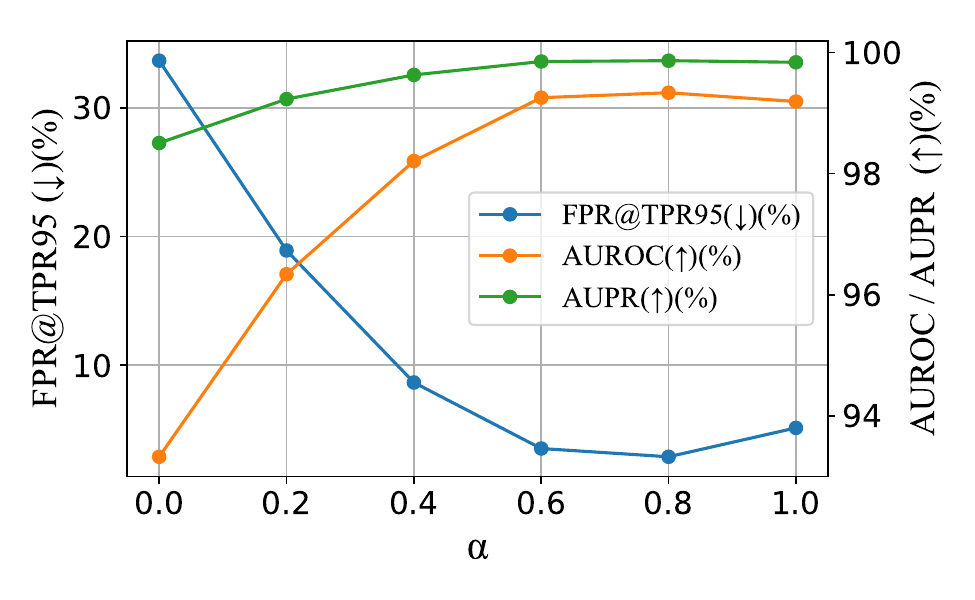}
 	\caption{Detection performance of ADD$^+$ (window size: 32) with respect to adjustment coefficient $\alpha$. Results are produced with a target model trained on CIFAR-100. ID: CIFAR-100 test set, OOD: CIFAR-10 training set.}
	\label{fig.alpha}
 \end{figure}

\textbf{Impact of feature adjustment.}
Based on the above results, we know that the vanilla ADD already achieves good detection ability for the four relative simple datasets with a window size of 32 or 64. However, its performance on CIFAR-100 is worse than others. 
Next we investigate whether employing the proposed feature adjustment helps to reduce the detection error on this dataset.
Fig.\ref{fig.alpha} shows the detection performance of ADD$^+$ in terms of three metrics with respect to the adjustment coefficient $\alpha$ on the CIFAR-100 model. The window size is fixed to 32. It is clear that we have a more accurate detection by introducing feature adjustment ($\alpha >0$). Specifically, the FPR@TPR95 reduces from more than 30\% to below 5\%, AUROC increases from about 93.5\% to above 99\%, and AUPR from 98.5\% to around 100\% when $\alpha = 0.8$. This means that with shift-adjustment, the detector also achieves competitive detection performance on CIFAR-100 even with a small window size.

\begin{figure*}[!t]
		\centering
		\subfloat[\scriptsize{MNIST (benign: USPS-7 and USPS-3)}]{
			\includegraphics[width=0.24\linewidth]{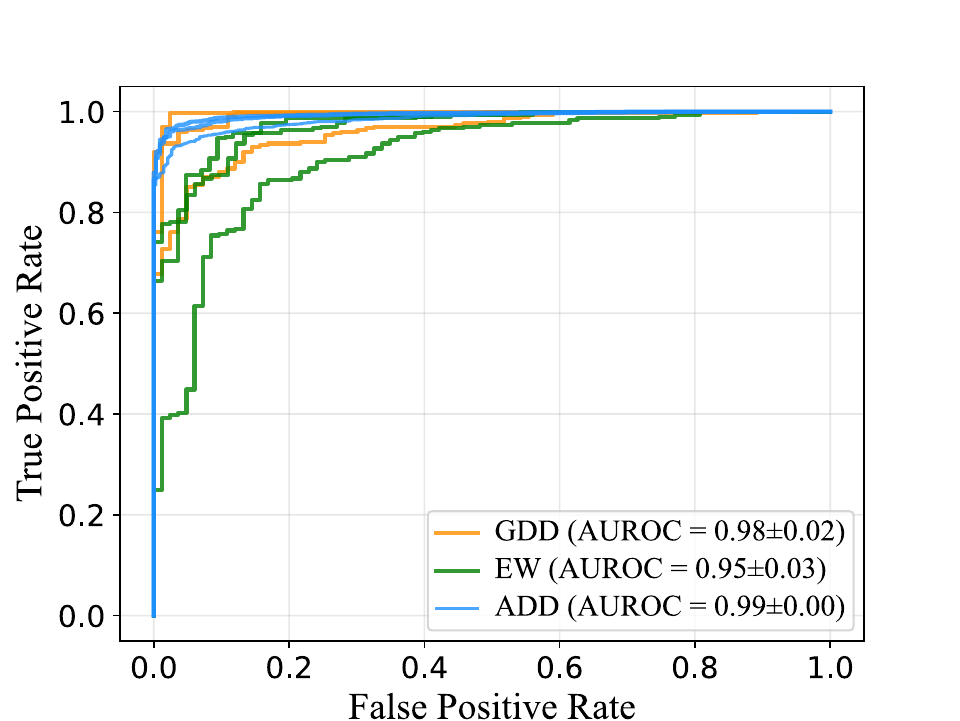}
			\includegraphics[width=0.24\linewidth]{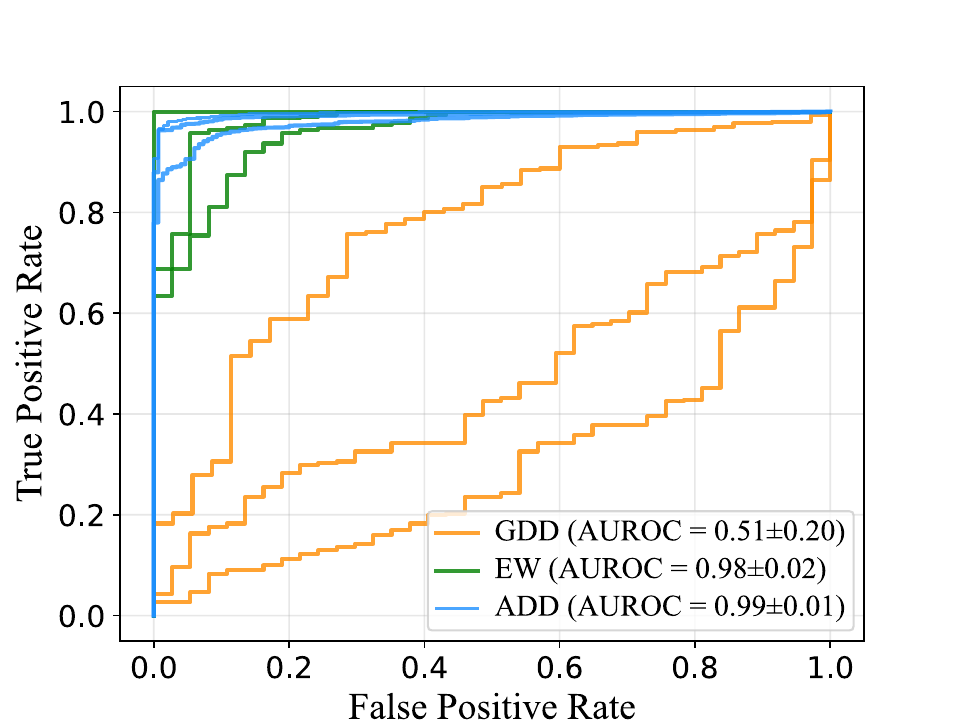} 		}
		\subfloat[\scriptsize{CIFAR-10 (benign: STL-7 and STL-3)}]{
			\includegraphics[width=0.24\linewidth]{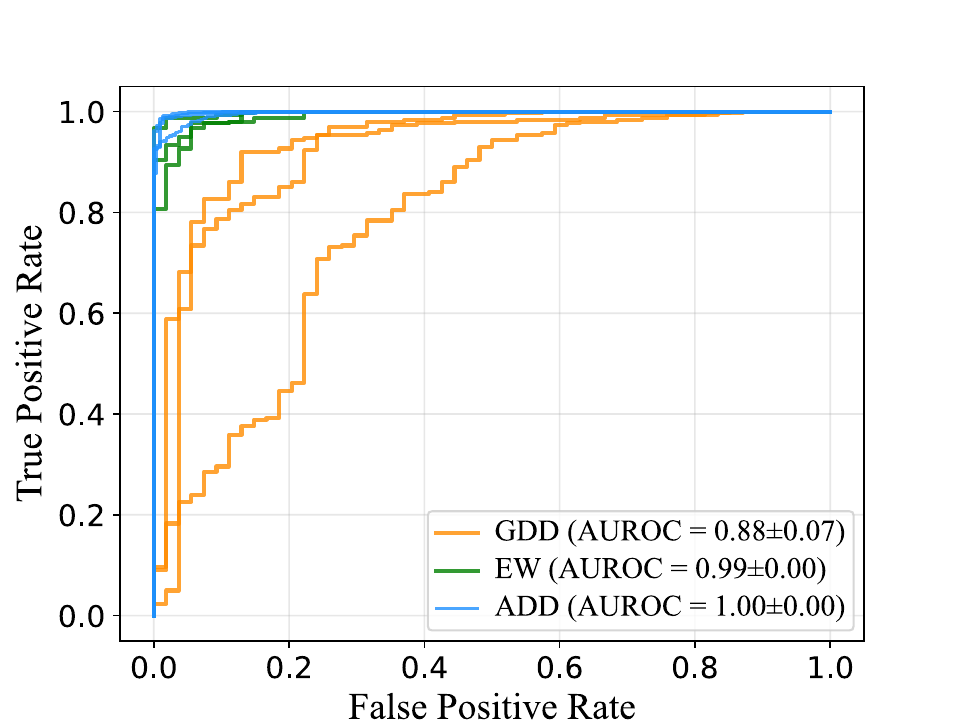}
			\includegraphics[width=0.24\linewidth]{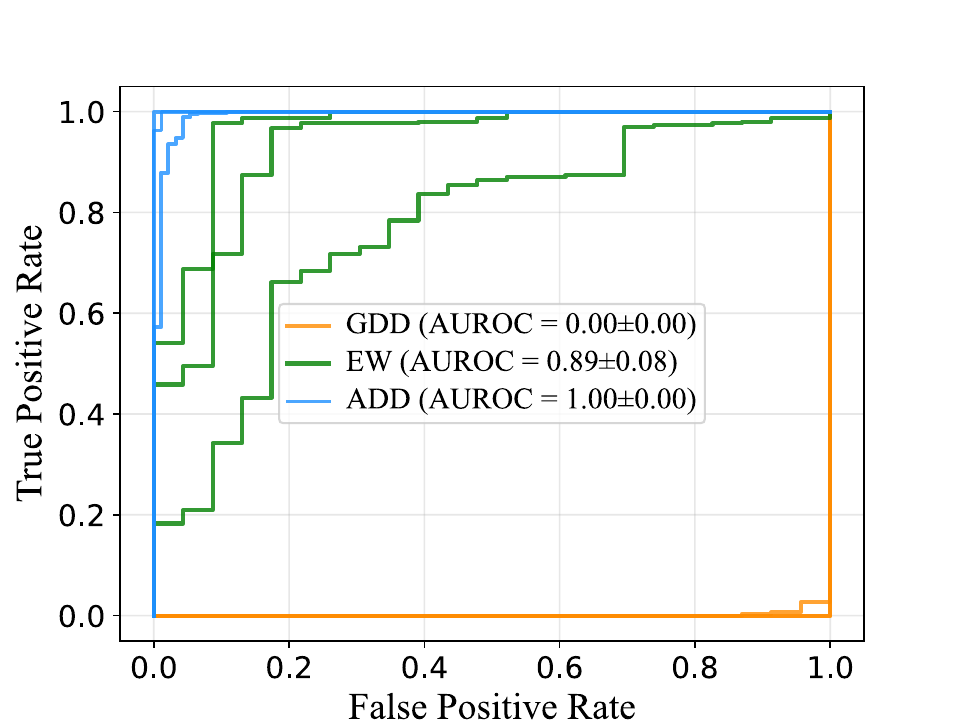} 	}	
			
		\caption{ROC curves of ADD and two simplified variants for MNIST (malicious: FashionMNIST) and CIFAR-10 (malicious: CIFAR-100). We repeat three times to generate the benign queries by randomly selecting a given number of classes. Results are obtained without feature adjustment. }
		\label{fig:ROC} 		
\end{figure*}

\textbf{Comparison with two simplified variants.}
Fig. \ref{fig:ROC} 	plots the ROC curves of the default ADD detector and its two simplified variants.  We observe that the default formulation is more robust to distribution shifts than the other two by maintaining high AUROC values in all the cases.  GDD is very sensitive to shifts caused by label inconsistency and performs poorly when benign queries only involves a small number of classes. This indicates the importance of measuring distribution distance in a class-conditional way. Weighting the class-wise distribution distance differently is also important to further improve the detection performance as shown by comparing EW and ADD.

\begin{table*}
    \caption{Comparison with existing defense methods against different attacks.}			\label{tab:results_knockoffnets}
		\centering
		\resizebox{1\textwidth}{!}{
			\begin{tabular}{ccccccccccccccc} 
				\toprule
				&\multirow{3}{*}{\textbf{Dataset}}  & &\multicolumn{12}{c}{\textbf{clone model acc.}(\%)}\\\cmidrule{5-15}
				&&\multicolumn{2}{c}{ \textbf{target model} (acc.\%)}&  \multicolumn{5}{c}{\textbf{KnockoffNets} (soft label)}&  \multicolumn{3}{c}{\textbf{DFME} (soft label) } & \multicolumn{3}{c}{\textbf{DFMS} (hard label)} \\ \cmidrule(r){3-4}
                 \cmidrule(r){5-9} \cmidrule(r){10-12} \cmidrule(r){13-15} 
				&                          & \textbf{\textit{undef}} &\textbf{\textit{D-ADD}} (acc. drop) & \textbf{\textit{undef}}&\textbf{\textit{AM}}  & \textbf{\textit{EDM}} &\textbf{\textit{CIP}}&\textbf{\textit{D-ADD}}   &\textbf{\textit{undef}} & \textbf{\textit{MeCo}} & \textbf{\textit{D-ADD}} &\textbf{\textit{undef}} & \textbf{\textit{MeCo}} & \textbf{\textit{D-ADD}}  \\ \midrule
				& MNIST        & 99.25 & 98.97  (-0.28) & 97.23   & 83.5   & 51.34  & 87.23 & \textbf{10.18}  & 98.85 & 79.98  & \textbf{13.15} & 97.95  & 98.39 & \textbf{12.72} \\
				& FashionMNIST & 90.51 & 90.51  (-0.00) & 48.67   & 27.5   & 15.28  & 31.42 & \textbf{10.00}  & 87.51 & 70.20  & \textbf{10.72} & 73.78  & 74.51 & \textbf{10.95} \\
				& CIFAR-10     & 94.27 & 94.27  (-0.00) & 87.11   & 77.2   & 68.50  & 75.59 & \textbf{12.42}  & 91.12 & 77.25  & \textbf{10.49 }& 89.01  & 80.51 & \textbf{10.24} \\
				& CIFAR-100    & 74.53 & 74.16  (-0.50) & 49.31   & 51.0   & 41.16  & 32.77 & \textbf{3.15}   & 52.27 &22.12   & \textbf{1.05}  & 68.71  & 65.85 & \textbf{2.85}  \\
				& Flowers-17   & 88.75 & 88.75  (-0.00) & 82.35   & 27.2   & 30.15  & 12.93 & \textbf{6.88}   & 64.12 &27.06   & \textbf{5.88}  & 67.05  & 60.48 & \textbf{5.61}  \\ \bottomrule
			\end{tabular}
		}
\end{table*}

\subsection{Experiments-II: Model Stealing \label{sec:exp_stealing}} 

\textbf{Attacks.} We adopt the same architecture for the clone model as the target model, assuming the attacker knows the target model's architecture. This setting makes  model stealing becomes easier while defense more challenging.  
It has been reported in ~\cite{kariyappa2021protecting} that JBDA-type attacks have low clone model accuracy as they are designed to assist adversarial-example generation instead of functionality replication.
Therefore, we here only consider KnockoffNets ~\cite{orekondy2019knockoff} and two data-free attack methods: DFME ~\cite{DFME_truong2021} and DFMS~\cite{DFMS_sanyal2022}.  

\textbf{Malicious queries for KnockoffNets.}
For each target model, we follow (Kariyappa and Qureshi 2020) to use a different dataset (3rd column) as the surrogate data to simulate malicious queries in KnockoffNets attack.
\begin{table}[h]
	\centering
    	\caption{ Architectures of target models and surrogate set used as malicious queries}
	\resizebox{\columnwidth}{!}{
		\begin{tabular}{ccccc}
			\toprule
			\textbf{architecture}                       & \textbf{training data}   & \textbf{malicious queries (number)}     \\ \midrule
			\multirow{2}{*}{Conv3}	                     & MNIST      &FashionMNIST (60,000)  \\
			& FashionMNIST &  MNIST (60,000) \\\midrule
			\multirow{2}{*}{WRN-16-4} & CIFAR-10 &CIFAR-100  (50,000)   \\
			& CIFAR-100   &CIFAR-10  (50,000) \\\midrule
			ResNet-18               & Flowers-17 &Indoor-67  (15,620) \\ \bottomrule
			\label{t1}
		\end{tabular}
	}
	\label{tab:architecture}
\end{table}

\textbf{Hyperparameter settings.} For the proposed D-ADD defense, we set the training accuracy dropping ratio $\gamma$ to $10^{-4}$ and $\alpha = 0.5$. The sliding window size $N$ is set to be 32 for CIFRA-100 and Flower-17, and 16 for the other three. Without loss of generality, we consider that each user query contains one image. 
The clone model is trained using the SGD optimizer for 50 epochs with a cosine annealing schedule and an initial learning rate of 0.1. \par
\textbf{Evaluation.} The performance of target and clone classification models are evaluated using the most widely used Top-1 classification accuracy on the corresponding testing data. Following existing studies, we evaluate a defense approach against model stealing with respect to its impact on the performance of the target model and the clone model. Specifically, a smaller drop in accuracy of the target model for benign queries indicates higher \underline{\textit{utility}}, while a larger performance drop of the clone model in the target classification task demonstrates stronger \underline{\textit{protection}}.

\textbf{Compared defense approaches.} We compare D-ADD with state-of-the-art defense approaches, including AM\cite{kariyappa2020defending},
EDM~\cite{kariyappa2021protecting}, CIP~\cite{CIP@TIFS23}, and MeCo~\cite{defenseDFME23}. The first three deal with KnockoffNets while the last one is designed for data-free attacks. Specifically, 
\begin{itemize}
	\item AM trains an OOD detector together with the target model and learns an auxiliary ``misinformation model" for returning poisoned predictions to malicious queries.
	\item EDM uses an ensemble of models jointly trained to give dissimilar predictions for OOD samples to disrupt effective cloning.
	\item CIP uses pre-collected malicious data to enlarge the discrepancy between the energy distribution of malicious and benign queries.
    \item MeCo is particularly designed for defending against data-free modal stealing by leveraging the randomized input strategy to disrupt the attacker's knowledge distillation.
\end{itemize}\par

\textbf{Comparison of defense performance against stealing.}
Table \ref{tab:results_knockoffnets} compares the performance of different defense methods against three attacks. It is seen from the results of undefended clone model that KnockoffNets performs well with proper surrogate data. DFME and DFMS achieve stable performance without using any natural-image queries but at a larger query cost compared to KnockoffNets. 
	
The clone model accuracy drops when defense is applied. Compared to other approaches, D-ADD achieves significantly better defense capability.  In all cases, the clone model's performance approaches the level of a random classifier when D-ADD is employed. While working with hard labels enables the DFMS attack to perfectly evade detection-free defense approaches including MeCo, the protection of D-ADD remains the same like soft label settings.

As seen from the second and third columns, performance of the target models is largely preserved when D-ADD is applied, with less than 0.1\% accuracy drop for the worst case.

\subsection{Experiments-III: Adaptive Attacks}  \label{sec:exp_adaptiveAttack}

\textbf{Settings and evaluation.}
We consider that the attacker has a small number of in-distribution (ID) samples as  benign-looking images, which are uniformly mixed with surrogate samples, so that the percentage of ID samples in a window is always close to the overall mix-up ratio. To evaluate the robustness of D-ADD, we compare the performance of the models $mix\_clone$ and $id\_clone$. 
The $mix\_clone$ model is trained using all the queries, while the $id\_clone$ model is trained using the same set of ID queries only. The D-ADD defense is considered robust against this kind of adaptive attack if $mix\_clone$ performs better than $id\_clone$
with no defense but worse when D-ADD is
employed, i.e., the following holds in term of accuracy:
\[
undef(mix\_clone) >  id\_clone > \text{D-ADD}(mix\_clone)
\]
The intuition is that the attacker would not pay extra cost for collecting and querying with OOD surrogate samples unless there is a significant accuracy gain.
	
To simulate this scenario, we performed an adaptive KnockoffNets attack by mixing queries in each sliding window with a certain mix-up ratio. The window size for each dataset is set in the same way as in the previous experiment. Generally, the more unique benign-looking samples are used as queries, the more likely that $id\_clone$ performs better than $mix\_clone$. Based on the assumption of data limitation as well as the above condition, we limited the percentage of benign-looking samples to be less than 10\% of surrogate queries for CIFAR-10, CIFAR-100 and Flower-17, and less than 1\% for other two simple tasks.

\textbf{Results.}
Table \ref{tab:apdative} displays the test accuracy of clone models $mix\_clone$ and $id\_clone$. It clearly shows that $mix\_clone$ outperforms $id\_clone$ in all cases when the target model is not protected with any defense technique. However, when D-ADD is employed, the accuracy of $mix\_clone$ becomes worse than $id\_clone$. According to the evaluation criterion given earlier in this section, D-ADD effectively resists mixing-based adaptive attacks.

\begin{table}[]
		\centering
        \caption{Robustness of D-ADD to adaptive attack.}		\label{tab:apdative}
		\resizebox{\columnwidth}{!}{
			\begin{tabular}{ccccc}
				\toprule
				\multirow{2}{*}{\textbf{Dataset}}&\multirow{2}{*}{\textbf{No. ID queries}} &   \multicolumn{2}{c}{\textbf{mix\_clone} (acc.\%) }& \multirow{2}{*}{\textbf{id\_clone }(acc.\%) }\\ \cmidrule{3-4} 
				&  & \textbf{\textit{undef}} &\textbf{\textit{D-ADD}} & \\			
				\midrule
				\multirow{2}{*}{MNIST}   
				& 60      & 98.00  & 10.58      &    55.27      \\ 
				& 600     & 98.72  & 10.81      &    91.94      \\ 	\midrule
				\multirow{2}{*}{FashionMNIST}
				& 60      & 63.58  & 11.03      &    19.35      \\ 
				& 600     & 81.00  & 13.66      &    73.88      \\ 
				\midrule
				\multirow{2}{*}{CIFAR-10} 
				& 500     & 87.38   & 11.01     &    36.28      \\ 
				& 5000    & 89.44   & 27.46     &    70.48      \\	\midrule
				\multirow{2}{*}{CIFAR-100} 
				& 500     & 52.32   &  5.14     &     9.67       \\ 
				& 5000    & 61.72   & 14.03     &    33.16      \\\midrule
				\multirow{2}{*}{Flower-17} 
				& 100     & 80.59   & 11.45     &    31.55      \\ 
				& 500     & 85.88   & 12.35     &    51.76      \\
				\bottomrule
				
			\end{tabular}
		}
	\end{table}
\section{Limitation}
Sybil attacks can be a threat to account-based defense methods
like D-ADD and PRADA as attackers can spread their queries
across different accounts. However, D-ADD is less vulnerable to
sybil attacks compared to other methods because it works with a
small query window rather than all historical queries. To generate
good clone models, KnockoffNets requires 6 million queries, while
data-free attacks require 20 million. To effectively evade D-ADD,
which operates with a window of size 64 or less, attackers need to create
thousands of accounts, incurring additional setup costs. It’s possible to trade off some defense for better robustness to sybil attacks
by reducing the window size. Additionally, service providers commonly limit the number of accounts created by each user using
other network-level means.

\section{Conclusion} \label{sec:conclusion}

Although real-time defense offers timely protection for the target model against stealing attacks, the key to ensuring robust protection lies in distinguishing malicious queries from those submitted by benign users, a challenging task in practical scenarios.
We have investigated a new direction to design malicious detectors by leveraging local dependency between nearby queries in the stream from the same account,  and instantiated this idea with a feature space nonparametric detector based on weighted class-wise distribution discrepancy. 
We provide analytical discussion and experimental evidence to show that the proposed defense achieves competitive protection with preserved utility under challenging settings, making it an important complement to research on the model stealing defense.  However, like other approaches, the detection and robustness of our approach is only ensured under certain circumstances.  More efforts are needed in the future to make it more suitable for practical use.  It is also important to consider integrating it with other defensive means to handle various risks in complex real-world scenarios, such as advanced adaptive attacks and sybil attacks.

\bibliography{sfd1}
\bibliographystyle{IEEEtran}

\end{document}